\documentclass[twocolumn]{aastex63}

\usepackage{tabularx}

\received{September 30, 2020}
\revised{November 19, 2020}
\accepted{December 3, 2020}
\submitjournal{ApJ}

\shorttitle{A new $\Delta$S calibration for RR Lyrae variables}
\shortauthors{Crestani et al.}

\graphicspath{{./}{}}

\begin{document}

\title{On the Use of Field RR Lyrae as Galactic Probes. II.
A new $\Delta$S calibration to estimate their metallicity
\footnote{Based on observations obtained with the du Pont telescope at Las Campanas Observatory, operated by Carnegie Institution for Science.
Based in part on data collected at Subaru Telescope, which is operated by the National Astronomical Observatory of Japan.
Based partly on data obtained with the STELLA robotic telescopes in Tenerife, an AIP facility jointly operated by AIP and IAC.
Some of the observations reported in this paper were obtained with the Southern African Large Telescope (SALT).
Based on observations made with the Italian Telescopio Nazionale Galileo (TNG) operated on the island of La Palma by the Fundación Galileo Galilei of the INAF (Istituto Nazionale di Astrofisica) at the Spanish Observatorio del Roque de los Muchachos of the Instituto de Astrofisica de Canarias.
Based on observations collected at the European Organisation for Astronomical Research in the Southern Hemisphere.
}
}

\correspondingauthor{J. Crestani}
\email{juliana.crestani@uniroma1.it}

\author{J. Crestani}
\affiliation{Dipartimento di Fisica, Universit\`a di Roma Tor Vergata, via della Ricerca Scientifica 1, 00133 Roma, Italy}
\affiliation{INAF -- Osservatorio Astronomico di Roma, via Frascati 33, 00078 Monte Porzio Catone, Italy}
\affiliation{Departamento de Astronomia, Universidade Federal do Rio Grande do Sul, Av. Bento Gon\c{c}alves 6500, Porto Alegre 91501-970, Brazil}

\author{M. Fabrizio}
\affiliation{INAF -- Osservatorio Astronomico di Roma, via Frascati 33, 00078 Monte Porzio Catone, Italy}
\affiliation{Space Science Data Center -- ASI, via del Politecnico snc, 00133 Roma, Italy}

\author{V.~F. Braga}
\affiliation{INAF -- Osservatorio Astronomico di Roma, via Frascati 33, 00078 Monte Porzio Catone, Italy}
\affiliation{Space Science Data Center -- ASI, via del Politecnico snc, 00133 Roma, Italy}

\author{C. Sneden}
\affiliation{Department of Astronomy and McDonald Observatory, The University of Texas, Austin, TX 78712, USA}

\author{G. Preston}
\affiliation{The Observatories of the Carnegie Institution for Science, 813 Santa Barbara St., Pasadena, CA 91101, USA}

\author{I. Ferraro}
\affiliation{INAF -- Osservatorio Astronomico di Roma, via Frascati 33, 00078 Monte Porzio Catone, Italy}

\author{G. Iannicola}
\affiliation{INAF -- Osservatorio Astronomico di Roma, via Frascati 33, 00078 Monte Porzio Catone, Italy}

\author{G. Bono}
\affiliation{Dipartimento di Fisica, Universit\`a di Roma Tor Vergata, via della Ricerca Scientifica 1, 00133 Roma, Italy}
\affiliation{INAF -- Osservatorio Astronomico di Roma, via Frascati 33, 00078 Monte Porzio Catone, Italy}

\author{A. Alves-Brito}
\affiliation{Departamento de Astronomia, Universidade Federal do Rio Grande do Sul, Av. Bento Gon\c{c}alves 6500, Porto Alegre 91501-970, Brazil}

\author{M. Nonino}
\affiliation{INAF -- Osservatorio Astronomico di Trieste, Via G. B. Tiepolo 11, 34143 Trieste, Italy}

\author{V. D'Orazi}
\affiliation{INAF -- Osservatorio Astronomico di Padova, vicolo dell’Osservatorio 5, 35122, Padova, Italy}
\affiliation{School of Physics and Astronomy, Monash University, Clayton, VIC 3800, Melbourne, Australia}

\author{L. Inno}
\affiliation{INAF -- Osservatorio Astronomico di Capodimonte, Salita Moiariello 16, 80131 Napoli, Italy}
\affiliation{Universit\`a Parthenope di Napoli, Science and Technology Department, CDN IC4, 80143 Naples, Italy}

\author{M. Monelli}
\affiliation{Instituto de Astrof\'isica de Canarias, Calle Via Lactea s/n, E38205 La Laguna, Tenerife, Spain}

\author{J. Storm}
\affiliation{Leibniz-Institut f\"ur Astrophysik Potsdam (AIP), An der Sternwarte 16, D-14482 Potsdam, Germany}

\author{G. Altavilla}
\affiliation{INAF -- Osservatorio Astronomico di Roma, via Frascati 33, 00078 Monte Porzio Catone, Italy}
\affiliation{Space Science Data Center -- ASI, via del Politecnico snc, 00133 Roma, Italy}

\author{B. Chaboyer}
\affiliation{Department of Physics and Astronomy, Dartmouth College, Hanover, NH 03784, USA}

\author{M. Dall'Ora}
\affiliation{INAF -- Osservatorio Astronomico di Capodimonte, Salita Moiariello 16, 80131 Napoli, Italy}

\author{G. Fiorentino}
\affiliation{INAF -- Osservatorio Astronomico di Roma, via Frascati 33, 00078 Monte Porzio Catone, Italy}

\author{C. Gilligan}
\affiliation{Department of Physics and Astronomy, Dartmouth College, Hanover, NH 03784, USA}

\author{E.~K. Grebel} 
\affiliation{Astronomisches Rechen-Instit\"ut, Zentrum f\"ur Astronomie der Universit\"at Heidelberg, M\"onchhofstr. 12-14, D-69120
Heidelberg, Germany}

\author{H. Lala}
\affiliation{Astronomisches Rechen-Instit\"ut, Zentrum f\"ur Astronomie der Universit\"at Heidelberg, M\"onchhofstr. 12-14, D-69120
Heidelberg, Germany}


\author{B. Lemasle}
\affiliation{Astronomisches Rechen-Instit\"ut, Zentrum f\"ur Astronomie der Universit\"at Heidelberg, M\"onchhofstr. 12-14, D-69120
Heidelberg, Germany}

\author{M. Marengo}
\affiliation{Department of Physics and Astronomy, Iowa State University, Ames, IA 50011, USA}

\author{S. Marinoni}
\affiliation{INAF -- Osservatorio Astronomico di Roma, via Frascati 33, 00078 Monte Porzio Catone, Italy}
\affiliation{Space Science Data Center -- ASI, via del Politecnico snc, 00133 Roma, Italy}

\author{P.~M. Marrese}
\affiliation{INAF -- Osservatorio Astronomico di Roma, via Frascati 33, 00078 Monte Porzio Catone, Italy}
\affiliation{Space Science Data Center -- ASI, via del Politecnico snc, 00133 Roma, Italy}

\author{C.~E. Mart\'inez-V\'azquez}
\affiliation{Cerro Tololo Inter-American Observatory, NSF's National Optical-Infrared Astronomy Research Laboratory, Casilla
603, La Serena, Chile}

\author{N. Matsunaga}
\affiliation{Department of Astronomy, The University of Tokyo, 7-3-1 Hongo, Bunkyo-ku, Tokyo 113-0033, Japan} 

\author{J.~P. Mullen}
\affiliation{Department of Physics and Astronomy, Iowa State University, Ames, IA 50011, USA}

\author{J. Neeley}
\affiliation{Department of Physics, Florida Atlantic University, 777 Glades Rd, Boca Raton, FL 33431 USA}

\author{Z. Prudil}
\affiliation{Astronomisches Rechen-Instit\"ut, Zentrum f\"ur Astronomie der Universit\"at Heidelberg, M\"onchhofstr. 12-14, D-69120
Heidelberg, Germany}

\author{R. da Silva}
\affiliation{INAF -- Osservatorio Astronomico di Roma, via Frascati 33, 00078 Monte Porzio Catone, Italy}
\affiliation{Space Science Data Center -- ASI, via del Politecnico snc, 00133 Roma, Italy}

\author{P.~B. Stetson}
\affiliation{Herzberg Astronomy and Astrophysics, National Research Council, 5071 West Saanich Road, Victoria, British
Columbia V9E 2E7, Canada}

\author{F. Th\'evenin}
\affiliation{Universit\'e de Nice Sophia-antipolis, CNRS, Observatoire de la C\^ote d'Azur, Laboratoire Lagrange, BP 4229, F-06304 Nice, France}

\author{E. Valenti}
\affiliation{European Southern Observatory, Karl-Schwarzschild-Str. 2, 85748 Garching bei Munchen, Germany}
\affiliation{Excellence Cluster ORIGINS, Boltzmann\--Stra\ss e 2, D\--85748 Garching bei M\"{u}nchen, Germany}

\author{A. Walker}
\affiliation{Cerro Tololo Inter-American Observatory, NSF's National Optical-Infrared Astronomy Research Laboratory, Casilla
603, La Serena, Chile}

\author{M. Zoccali}
\affiliation{Instituto de Astrof\'sica, Facultad de F\'sica, Pontificia Universidad Cat\'lica de Chile, Av. Vicu\~na Mackenna 4860, Santiago, Chile}

\begin{abstract}

We performed the largest and most homogeneous spectroscopic survey of field 
RR Lyraes (RRLs). We secured $\approx$6,300 high resolution 
(HR, R$\sim$35,000) spectra for 143 RRLs (111 fundamental, 
RRab; 32 first overtone, RRc). The atmospheric parameters were estimated by using 
the traditional approach and the iron abundances were measured by using an 
LTE line analysis. The resulting iron distribution shows a well defined 
metal-rich tail approaching solar iron abundance. 
This suggests that field RRLs experienced a complex chemical enrichment in the 
early halo formation. We used these data to develop a new 
calibration of the $\Delta$S method. This diagnostic, based on the equivalent 
widths of CaII K and three Balmer (H$_{\delta,\gamma,\beta}$) lines, 
traces the metallicity of RRLs. For the first time the new empirical calibration: 
i) includes spectra collected over the entire pulsation cycle; 
ii) includes RRc variables; 
iii) relies on spectroscopic calibrators covering more than three dex in iron abundance; 
iv) provides independent calibrations based on one/two/three Balmer lines. 
The new calibrations were applied to a dataset of both SEGUE-SDSS and degraded 
HR spectra totalling 6,451 low resolution (LR, R$\sim$2,000) spectra for 
5,001 RRLs (3,439 RRab, 1,562 RRc). This resulted in an iron distribution with 
a median\footnote{We employ the letter $\eta$ in this work to refer to the median values
of whatever quantity is being discussed, e.g. the median of a distribution, the median 
of the residuals in consideration.} $\eta$=-1.55$\pm$0.01 and $\sigma$=0.51 dex, 
in good agreement 
with literature values. We also found that RRc are  0.10 dex more 
metal-poor than RRab variables, and have a distribution with a smoother 
metal-poor tail. This finding supports theoretical prescriptions suggesting 
a steady decrease in the RRc number when moving from metal-poor to 
metal-rich stellar environments.

\end{abstract}

\section{Introduction} \label{sec:intro}
The RR Lyrae (RRL) stars are radially pulsating, core helium burning stars. The fundamental mode and 
the first overtone pulsators are categorized as RRab and RRc, respectively, with the RRd being a small 
population of double mode pulsators. Being evolved low mass stars, they are ubiquitous in a 
variety of environments and cover a wide metallicity range ([Fe/H] $\approx$ -2.5 to solar). Interestingly,
the RRL obey period-luminosity relations in near-infrared bands 
\citep[e.g.][]{1986MNRAS.220..279L,2003MNRAS.344.1097B,2018AJ....155..137B, 2019ApJ...870..115B}. These 
relations make them ideal standard candles, like Cepheids, but with the unique characteristics of tracing 
necessarily old populations ($\geq$ 10 Gyr) and being very well studied \citep[e.g.][]{2015ApJ...808...50M}. 
Indeed, Classical Cepheids trace exclusively young populations, while Type II Cepheids are old but scarce
and still poorly understood \citep{2020arXiv200906985B}.

The absolute visual magnitude metallicity relation \citep{2006ARA&A..44...93S}
and the period-luminosity relations of the RRLs have made them very popular 
distance indicators for decades. Field RRLs have also been the focus of several 
spectroscopic studies \citep[e.g.][hereafter Layden94]{1959ApJ...130..507P,
1976ApJ...210..120B,1979AJ.....84..993B,1985ApJ...289..310S,1994AJ....108.1016L},
aimed at investigating the early chemical enrichment of both the Halo and the 
Bulge together with their three-dimensional spatial structure.
The question that immediately arises is whether chemical abundance 
determinations for these stars are reliable across their whole pulsation cycle. 
After all, even though their short (0.2 to 1 day) periods make their light 
curves relatively easy to obtain, the resulting rapid atmospheric changes 
could make detailed chemical abundance investigations very difficult,
in particular for fainter objects. 
In the past, this difficulty was two-fold: first, there was the challenge 
of observing the star for a sufficiently long  time in order to obtain a high 
signal-to-noise ratio (SNR), but not for so long as to cause the smearing 
of spectral lines due to the varying velocity of the atmospheric layer where 
the line is formed. Second, it was unclear whether the measured  atmospheric 
metallicity of the RRL remained constant across the pulsation cycle.

The first issue was solved by the development of larger telescopes and increased access to them. Even 2 meter class
telescopes have great success in obtaining a good SNR in high dispersion spectrographs for the brighter targets.
With 8 m class telescopes, exposures as short as a few minutes 
deliver high quality spectra for medium to bright targets, allowing a very detailed study of the atmospheric 
phenomena in these stars \citep{2008A&A...491..537C,2016A&A...587A.134G}. As for the second issue, the
breakthrough came with the works of \citet{2011ApJS..197...29F} for the RRab and \citet{2017ApJ...848...68S} for 
the RRc, both of which demonstrated that the observed metallicity of RRL remain constant within uncertainties 
for the entire pulsation cycle. The same result has been found by \citet{2019ApJ...881..104M} for
a large sample of RRL in the $\omega$ Centauri globular cluster.

These developments confirmed that coherent chemical abundance results can be recovered during the whole pulsation 
cycle, thus opening the door for random phase observations. This efficient approach to observations caused a
significant increase in the number of RRL with spectroscopic studies. Indeed, with the nonlinear effects caused by shockwaves
being restricted to a narrow phase window, a series of observations of any given RRL can easily 
provide perfectly useable data. However, high resolution spectroscopy is still a resource-consuming endeavour. 
A single RRL is capable of constraining both chemical abundance and distance, yet a detailed study of a stellar population
requires a large number of such observations. It is still unviable to acquire high dispersion spectra for objects beyond the 
Milky Way, or obtain and analyse a very large number of such spectra to perform detailed 
studies of the Galactic halo. Both stellar distance and sample size are strong constraints on the applicability of high 
resolution spectroscopy.

Originally, the $\Delta$S method consisted in deriving two spectral types for 
a given RRL, one only considering the Balmer series features (H$_{\beta}$, H$_{\gamma}$, 
H$_{\delta}$), and one only the CaII K line. The difference in spectral type 
between these two indicators was then associated to a difference in 
metallicity \citep{1959ApJ...130..507P,1975ApJ...200...68B}. Following this 
working hypothesis, the studies of \citet{1975ApJ...201L..71F}, 
\citet{1991ApJ...367..528S}, and \citet{1994AJ....108.1016L} 
demonstrated that the equivalent widths of the CaII K and three of the Balmer 
series features can also be associated to the metallicity of a given RRL without 
the use of the spectral type. These features are extremely strong and can be 
reliably measured at very low spectral resolutions. This means that, once 
a calibration using high dispersion spectroscopy is derived, low resolution 
measurements are enough to provide metallicity estimates. This equivalent width 
approach is the basis of the $\Delta $S method as it is discussed in the 
present work.

The $\Delta $S method is behind the majority of metallicity estimates of RRL stars in the literature directly, 
via its application to low resolution spectra, or indirectly, when its results are used to 
calibrate other metallicity indicators such as the Fourier parameter decomposition method. This low resolution 
spectroscopic metallicity indicator is a balanced approach between the precision of high resolution 
spectroscopy and the efficiency of photometry. \citet{2012Ap&SS.341...89W} have adapted the $\Delta$S method to
use the infrared CaII line at 8,498 \AA{} covered by the \emph{Gaia} spectrograph \citep{2018A&A...616A...5C}.
Another related method using the
CaII K line and the Balmer series features was employed by \citet{2020ApJS..247...68L}, using synthetic spectra 
matching instead of equivalent widths.

The calibration provided by Layden94 is widely used in the literature \citep{2013MNRAS.435.3206D}, although later
ones exist \citep[e.g.][]{2004A&A...421..937G}. It was produced using 19 RRab stars at minimum light phases and 
applied a metallicity scale transformation based on seven globular clusters to provide values in the 
\citet{1984ApJS...55...45Z} scale. In this work, we employed the largest and most homogeneous high resolution 
spectroscopic sample of field RRL stars ever used in the literature to provide a brand new calibration of 
$\Delta $S method across the entire pulsation cycle for both RRab and RRc stars, including in the metal-rich and 
metal-poor tails that were previously poorly sampled. 

Our sample allows us to both determine the metallicities in high resolution in a homogeneous way, and the $\Delta$S 
measurements once the spectra were downgraded to low resolution. With this, this new calibration makes use of 
no metallicity scale transformations. As we have dozens of calibrating stars, anyone who wishes to perform new 
measurements in our system has several good reference targets to choose from. This means no equivalent width system 
transformations are necessary either. Moreover, we streamlined some features of the IDL code
{\textsc{EWIMH}}\footnote{The code and documentation can be found at 
\url{http://physics.bgsu.edu/~layden/ASTRO/DATA/EXPORT/EWIMH/ewimh.htm}}, one of several codes traditionally 
employed for the $\Delta $S measurements, to make full use of the much higher quality spectrographs currently in use. 
These changes are easy to implement in other codes for the same purpose.

This paper is organized as follows. In Sect.~\ref{ch:datasets} we describe the spectroscopic datasets, and in 
Sect.~\ref{ch:define_samples} how they are
organised into samples. In Sect.~\ref{ch:deltas_definition_measurements} we discuss the changes made to the definition 
of the $\Delta $S method and the changes on how the measurements are performed by the code. The calibration between the 
$\Delta $S index and [Fe/H] is provided in Sect.~\ref{ch:ds_feh_calibration}. In Sect.~\ref{ch:validation} we compare
our results with the literature in both high and low resolution. Finally, the calibration is applied to a large sample 
of field RRLs
including hundreds of RRc and one RRd in Sect.~\ref{ch:application}. A summary and final remarks are given in 
Sect.~\ref{ch:final_remarks}. Further discussion regarding the differences between the current calibration and
Layden94, and phase considerations are presented in Appendix~\ref{ch:appendix_ew_feh_phase}.

\section{Spectroscopic data sets} \label{ch:datasets}

To provide firm constraints on the metallicity distribution of the
Galactic halo, we adopted the same photometric catalog built in F19, 
the first paper of this series.
To take into account new optical RRL catalogs and surveys that have been
recently published the original photometric catalog was complemented
with four new data set provided by
Catalina \citep{2017MNRAS.469.3688D},
PANSTARRS \citep{2017AJ....153..204S},
ASA-SN \citep{2019MNRAS.485..961J} and
DECAM \citep{2019AJ....158...16S}.
The selection criteria are the same discussed in F19. The new final
photometric catalog includes  $\sim$179,000 RRLs and was the starting
point for collecting the spectroscopic dataset. A detailed description
of the construction of this catalog is provided in a companion paper
(Braga et al. 2020, in preparation).

\subsection{High resolution spectroscopic data set} \label{ch:hr_dataset}
We have collected a sample of 6,631 spectra for 266 stars (173 RRab, 92 RRc, 1 RRd, Tab.~\ref{tab:stars}) across the 
pulsation phase from multiple medium and high resolution spectrographs. The largest 
and most homogeneous subsample was collected with the echelle spectrograph at du Pont (Las Campanas Observatory) and 
includes 6,070 high resolution (R=35,000) spectra of 186 RRL (107 RRab, 79 RRc). It typically covers wavelengths from 
3,600 to 9,300 \AA{}. 

\begin{deluxetable*}{lcrrrrrlrr} 
\tablecaption{Identification, photometric characteristics, classification, and number of spectra for the CHR. \label{tab:stars}}
\tablewidth{\textwidth}
\tablehead{
\colhead{GaiaID} & \colhead{Star} & 
\colhead{RA$_{J2000}$} & \colhead{Dec$_{J2000}$} & \colhead{Vmag} & \colhead{Vamp} & \colhead{P} & 
\colhead{Class} & \colhead{Blazhko} & \colhead{N spectra} \\
\colhead{(DR2) } &\colhead{ } & 
\colhead{(deg)} &  \colhead{(deg)} & \colhead{(mag)} & \colhead{(mag)} & \colhead{(day)} & 
\colhead{ } & \colhead{ } & \colhead{ } \\
}
\startdata
4224859720193721856 & AA Aql & 309.5628 &  -2.8903 & 11.86 & 1.16 & 0.36 & RRab & ?    & 1   \\
2608819623000543744 & AA Aqr & 339.0161 & -10.0153 & 12.93 & 0.99 & 0.61 & RRab & ?    & 1   \\
1234729400256865664 & AE Boo & 221.8968 &  16.8453 & 10.62 & 0.45 & 0.31 & RRc  & ?    & 5   \\
3604450388616968576 & AM Vir & 200.8889 & -16.6663 & 11.45 & 0.69 & 0.62 & RRab & Yes  & 141 \\
1191510003353849472 & AN Ser & 238.3794 &  12.9611 & 10.92 & 1.01 & 0.52 & RRab & No   & 67  \\
\enddata
\tablecomments{Table \ref{tab:stars} is published in its entirety in the machine-readable format.
A portion is shown here for guidance regarding its form and content.
}
\end{deluxetable*}

\begin{figure}
\includegraphics[width=\columnwidth]{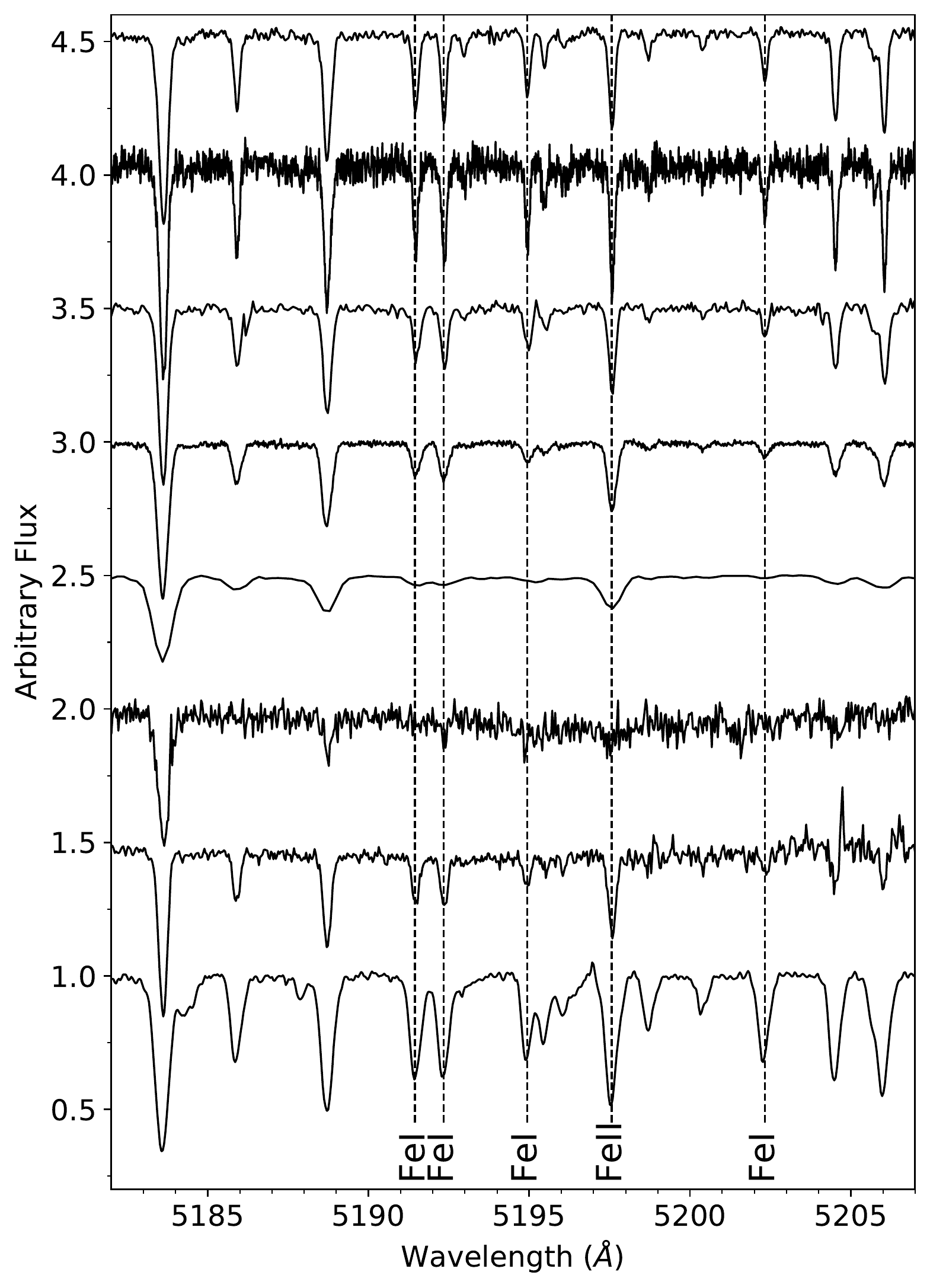}
\caption{Examples of high resolution spectra from all spectrographs used in this work. From top to bottom: FEROS,
HARPS, du Pont, UVES, X-shooter, Subaru HDS, STELLA, and SALT. The five first spectra are of V~Ind 
([Fe/H] $=$ -1.63 $\pm$ 0.03) near phase 0.40. They are followed by random phase spectra for 
X~Ari ([Fe/H] $=$ -2.59 $\pm$ 0.05), DH~Peg ([Fe/H] $=$ -1.37 $\pm$ 0.05), and
RW~Tra ([Fe/H] $=$ 0.13 $\pm$ 0.06). The dashed lines indicate the strongest iron lines present in this wavelength 
region.
\label{fig:hires_examples}}
\end{figure}

These data were complemented with 530 optical spectra from the ESO Archive. We included 
spectra from UVES and X-shooter at VLT (Cerro Paranal Observatory), HARPS at the 3.6m telescope (La Silla 
Observatory), and FEROS at the 2.2m telescope (La Silla Observatory). An additional 10 spectra were added from 
HARPS-N at the Telescopio Nazionale Galileo 
(Roque de Los Muchachos Observatory), 4 from HRS \citep{2014SPIE.9147E..6TC} at SALT
(South African Astronomical Observatory), 6 from the HDS \citep{2002PASJ...54..855N} 
at Subaru (National Astronomical 
Observatory of Japan) and 11 from the 
echelle spectrograph \citep{2012SPIE.8451E..0KW} at STELLA (Izana Observatory). 
Typical spectra for all spectrographs can be seen in Fig.~\ref{fig:hires_examples}.

\begin{figure}
\includegraphics[width=\columnwidth]{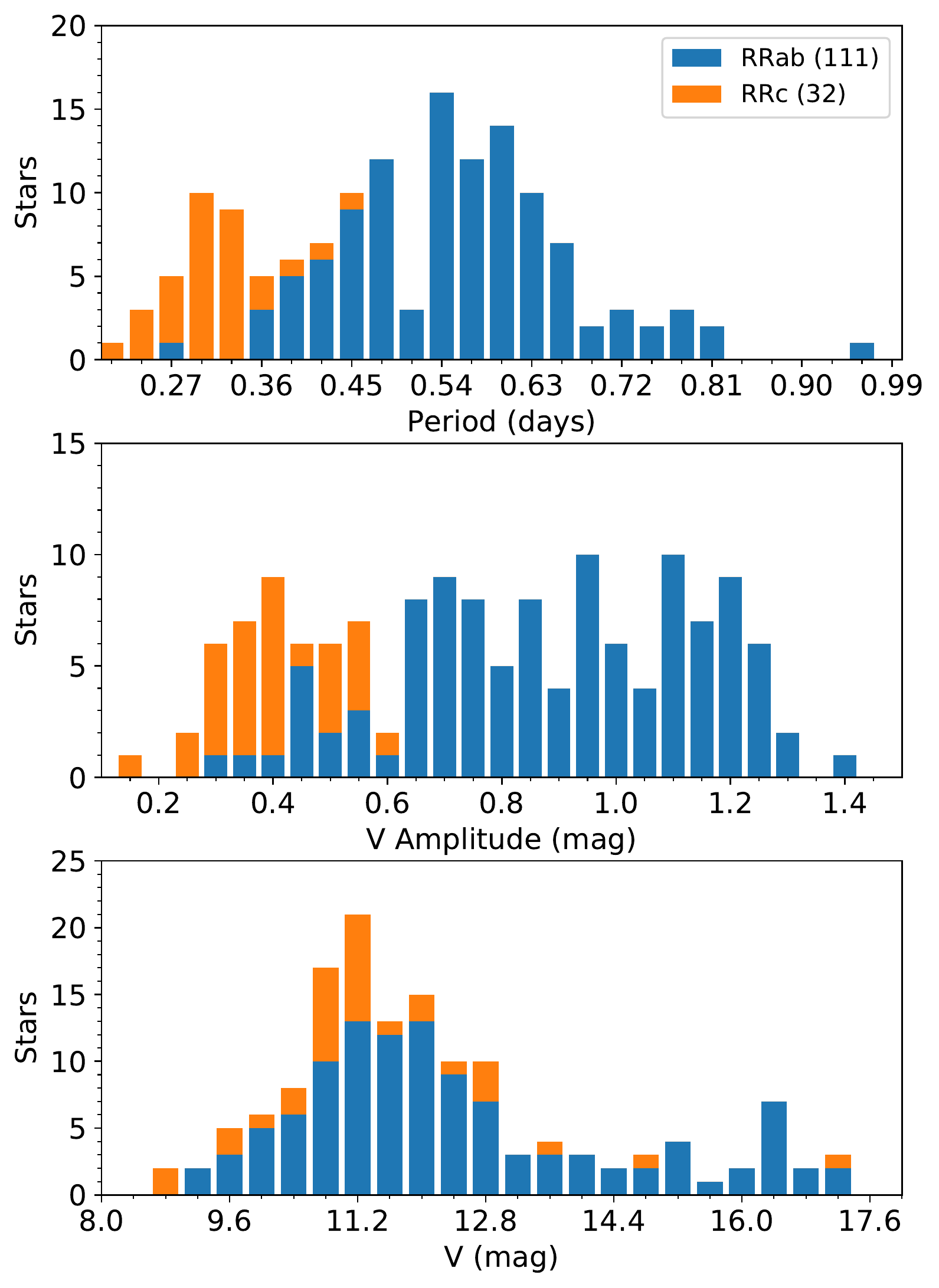}
\caption{Period (top), visual amplitude (middle), and apparent visual magnitude (bottom) distributions for the RRab 
(blue) and RRc (orange) in the calibrating sample.
\label{fig:P_Vamp_V_distribution}}
\end{figure}

The distribution of exposure times peaks at $\approx$400 seconds, with an average 434 seconds, and standard deviation
303 seconds. The reddening- and distance-independent photometric 
characteristics of the $\Delta$S calibrating sample, discussed in greater detail in Sect.~\ref{ch:define_samples},
 are shown in 
Fig.~\ref{fig:P_Vamp_V_distribution}. Visual magnitudes and amplitudes were derived from \emph{Gaia} G-band observations 
using the V-band transformation provided by \citet{2018A&A...616A...4E}. We also collected native V-band photometry 
from ASAS \citep{2002AcA....52..397P}, ASAS-SN \citep{2014ApJ...788...48S,2018MNRAS.477.3145J}, and Catalina 
\citep{2015MNRAS.446.2251T,2013ApJ...763...32D,2013ApJ...765..154D}. 

The normalizations and Doppler shift corrections were computed and applied using the National Optical Astronomy Observatory
libraries for {\textsc{IRAF}}\footnote{The legacy code is now maintained by the community on GitHub at 
\url{https://iraf-community.github.io/}} 
\citep[Image Reduction and Analysis Facility,][]{1993ASPC...52..173T}. In-depth information regarding the radial velocity 
studies and phasing applied to this same sample can be found in \citet{2020ApJ...896L..15B}. 

It is important to note that automatic continuum normalisation methods must be employed with caution. Using high 
order functions to fit the continuum may result in the CaII K and H lines being made significantly shallower than 
they truly are. This can occur in spectrographs where the continuum emission in each echelle order is not very 
smooth and requires higher order fitting functions. For the du Pont sample, due to its large size, we developed 
an algorithm that used the smoothest echelle orders neighboring the order where the line of interest was present. 
These neighbor orders, devoid of any strong absorption features, were used to determine the best continuum fit, 
which was then applied to the order of interest.

All high resolution spectra that displayed measurable CaII K and at least one Balmer hydrogen line (H$_{\delta}$, 
H$_{\gamma}$, H$_{\beta}$) were downgraded to R=2,000 for the measurement of the equivalent widths necessary for the
application of the $\Delta$S method (Sect.~\ref{ch:deltas_definition_measurements}). Spectra of stars for which we
have high resolution metallicity measurements from this work or 
\citet[][hereafter cited together as For+Chadid+Sneden]{2011ApJS..197...29F,2017ApJ...835..187C,2017ApJ...848...68S} form the calibrating sample. 
The rest were downgraded and rebinned to low resolution and included in the low resolution sample. Both samples ere
described in Sect.~\ref{ch:define_samples}.

\subsection{Low resolution spectroscopic dataset} \label{ch:lr_dataset}
We took advantage of the huge low resolution (R $\approx$ 2,000) 
spectroscopic dataset collected by the Sloan Extension for Galactic Exploration and 
Understanding Survey of the Sloan Digital Sky Survey \citep[SEGUE-SDSS,][]{2009AJ....137.4377Y}, 
with the 2.5m Sloan Foundation 
Telescope at the Apache Point Observatory. The selection criteria for the spectroscopic 
sample were already discussed in detail in the first paper of this series \citet[][hereafter F19]{2019ApJ...882..169F}. 
Note that the F19 investigation was only based on fundamental RRLs so, in the present work, the RRc are added 
for the first time. 

Most stars in the SEGUE-SDSS sample have only one or two exposures. To apply the new $\Delta$S 
calibration, we selected the spectra with SNR greater than 15, or that had SNR smaller than
15 but passed a visual inspection. The inspection was performed individually in order to assess 
the quality of the CaII K and Balmer lines. We ended up with a sample of 6,299 spectra for 4,883 
stars, of which 3,379 are RRab and 1,504 are RRc.


\section{RR Lyrae spectroscopic samples} \label{ch:define_samples}
The 143 stars (111 RRab, 32 RRc) for which we have HR metallicity measurements either from this work or 
from For+Chadid+Sneden, and also have $\Delta$S measurements in the new definition, were included in the calibrating
sample (CHR). Those, together with 67 stars (59 RRab, 8 RRc) with HR measurements in the literature that could 
be brought onto our scale, were included in the high resolution (HR) sample. Finally, the low resolution (LR) 
sample contains 5,001 stars (3,439 RRab, 1,562 RRc) for which high resolution estimates were not available. Each sample 
is discussed in greater detail below.

\subsection{High resolution calibrating sample} \label{ch:calibrating_sample}
High resolution metallicities were derived from 171 measurements of 111 calibrating stars (91 RRab, 20 RRc, 
Tab.~\ref{tab:individual_atmospheres}). Wherever possible, we analysed two 
exposures per star. In case there were no spectra with high enough SNR for abundance analysis (SNR $\gtrapprox$ 50), 
we stacked spectra obtained at the same pulsation phase. The method for the determination of the pulsation periods 
for the whole sample is described in \citet{2020ApJ...896L..15B}. Unlike the aforementioned work, however, we have 
defined the initial point of the pulsation cycle, i.e. phase zero, as the point where the magnitude 
along the decreasing branch of the light curve is equal to the mean magnitude of the variable \citep{2016AJ....152..170B}.
This initial point, also called the reference epoch, is arbitrary and only moves the light and velocity curves rigidly 
alongside the phase axis.

When stacking spectra, it is essential to make sure that all stacked spectra were 
collected at times where the stellar atmosphere had nearly the same thermodynamical configuration. Both observations 
\citep{2011ApJS..197...29F} and theoretical models \citep{1992MmSAI..63..357B} show that the amosphere of RRL stars
 undergo temperature 
changes as large as 1,000 K. The clearest sign of different atmospheric temperatures 
between two spectra is a difference between the strength of the absorption lines, as well as their profiles. At lower 
temperatures, metallic lines
are deeper and display larger EWs, while hydrogen lines are shallower and have smaller EWs. 

With this in mind, when
stacking spectra, we performed a phase selection as a first criterium. This ensures that the spectra were collected
when the star was in the same moment of its pulsation cycle. Afterwards, we visually inspected all spectra to be stacked 
to verify they showed similar equivalent widths and profiles for the same lines. Verifying the equivalent widths guarantees
that all combined spectra were collected roughly at the same pulsation phase. With this conservative approach, any 
precision concerns with the phasing do not affect the quality of the stacking. Furthermore, any possible issues with the 
rest frame velocity correction are detected at this point in case the central 
wavelengths of absorption features fail to align. This visual inspection avoids the artificial line smearing caused by 
stacking spectra where the Doppler shift has not been properly accounted for.

\begin{figure}
\includegraphics[width=\columnwidth]{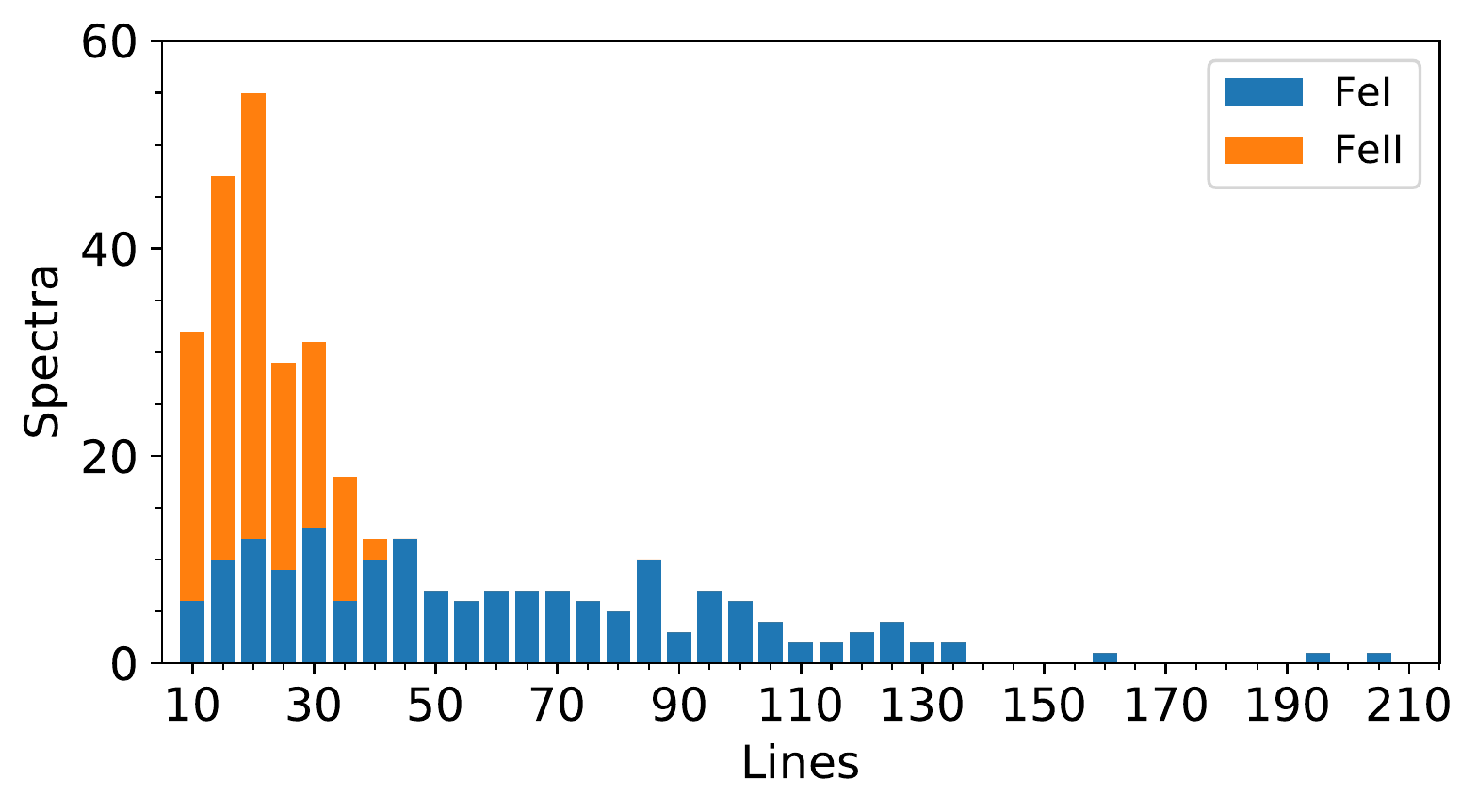}
\caption{Distribution of the number of adopted FeI (blue) and FeII (orange) lines per spectrum. A total of 174
spectra of 111 stars (91 RRab, 20 RRc) were used.
\label{fig:Nlines_distribution}}
\end{figure}

We built a line list of  of 415 FeI and 56 FeII lines with the most up-do-date transition parameters found in the 
literature, with a preference for laboratory values. The parameters and references are presented in 
Tab.~\ref{tab:linelist}. The lines with a reference that begin with the preffix \emph{NIST} had updated transition
parameters taken from the National Institute of Standards and Technology (NIST) Atomic Spectra Database
\footnote{Available at \url{https://www.nist.gov/pml/atomic-spectra-database}}.
The list was carefully cleaned of blended lines using both the \citet{1966sst..book.....M}
Solar spectrum atlas and synthetic spectra as a reference. The average number of useable lines in each spectrum was 
66 for FeI and 19 for FeII (Fig.~\ref{fig:Nlines_distribution}).

\begin{deluxetable}{lllrr} 
\tablecaption{List of FeI and FeII lines adopted in this work for the determination of atmospheric parameters.
\label{tab:linelist}}
\tablewidth{\columnwidth}
\tablehead{
\colhead{Wavelength} & \colhead{Species} & \colhead{EP} & \colhead{log(gf)} & \colhead{Source} \\
\colhead{(\AA{}) } & \colhead{ } & \colhead{(eV)} & \colhead{(dex)} & \colhead{ } \\
}
\startdata
3763.78  &  26.0  &  0.989  &  -0.220  &  OBR91	\\
3787.88  &  26.0  &  1.010  &  -0.840  &  OBR91	\\
3815.84  &  26.0  &  1.484  &   0.240  &  OBR91	\\	
3820.42  &  26.0  &  0.858  &   0.160  &  OBR91	\\	
3825.88  &  26.0  &  0.914  &  -0.020  &  OBR91	\\	
\enddata
\tablecomments{Table \ref{tab:linelist} is published in its entirety in the machine-readable format.
A portion is shown here for guidance regarding its form and content.
References: 
\citet[][BAR91]{1991AetA...248..315B}, \citet[][BAR94]{1994AetA...282.1014B}, \citet[][BEL17]{2017ApJ...848..125B},
\citet[][BLA79]{1979MNRAS.186..657B},  \citet[][BLA82]{1982MNRAS.199...43B},  \citet[][BLA86]{1986MNRAS.220..549B},
\citet[][BRI74]{1974ApJ...192..793B},  \citet[][DEN14]{2014ApJS..215...23D},  \citet[][KOC84]{1984PhST....8...84K},
\citet[][LAWLER]{2019ApJS..243...33D}, \citet[][MAY74]{1974AetAS...18..405M}, \citet[][MELBAR]{2009AetA...497..611M},
\citet[][OBR91]{1991JOSAB...8.1185O},  \citet[][RUF14]{2014MNRAS.441.3127R}. The preffix \emph{NIST} denotes the
line had its transition parameters updated by NIST.
}
\end{deluxetable}

Equivalent widths were measured manually using the {\textsc{IRAF}} task {\textsc{splot}} to perform a Gaussian fit
of the core of each line. There were no strong line asymmetries in any of the spectra used for this purpose.
 We performed an LTE line analysis using the 2019 
release of {\textsc{Moog}}\footnote{The code and documentation can be found at 
\url{https://www.as.utexas.edu/~chris/moog.html}} \citep{1973ApJ...184..839S} paired 
with an interpolated grid of $\alpha$-enhanced {\textsc{ATLAS9}} model atmospheres \citep{2003IAUS..210P.A20C}. Thus, 
we constrained the atmospheric parameters (effective temperature T$_{\text{eff}}$, surface gravity log(g), metallicity [Fe/H], 
and microturbulent velocity  $\xi_{\text{t}}$) using the traditional approach, i.e. iteratively changing their values until 
achieving excitation equilibrium of FeI lines, ionization equilibrium of FeI and FeII lines, and no trend between the 
abundances of FeI lines and their reduced equivalent widths.

\begin{deluxetable*}{lllcrrrrrrr} 
\tablecaption{Atmospheric parameters derived in this work for each individual measurement. \label{tab:individual_atmospheres}}
\tablewidth{\textwidth}
\tablehead{
\colhead{GaiaID} & \colhead{Star} & \colhead{Class} & \colhead{Spectrograph} &
\colhead{T$_{\text{eff}}$} & \colhead{log(g)} &  \colhead{$\xi_{\text{t}}$} & 
\colhead{[FeI/H]}  & \colhead{N$_{\text{FeI}}$} & \colhead{[FeII/H]} & \colhead{N$_{\text{FeII}}$}  \\
\colhead{(DR2)} & \colhead{ } & \colhead{ } & \colhead{ } & 
\colhead{(K)} & \colhead{(dex)} & \colhead{(km s$^{-1}$)} & 
\colhead{(dex)} & \colhead{ } & \colhead{(dex)} & \colhead{ } 
}
\startdata
4224859720193721856 & AA Aql & RRab & SALT   & 6610$\pm$65 & 2.70$\pm$0.01 & 2.52$\pm$0.04 & -0.34$\pm$0.24 & 206 & -0.34$\pm$0.22 & 39 \\
2608819623000543744 & AA Aqr & RRab & UVES   & 5840$\pm$80 & 1.52$\pm$0.06 & 3.51$\pm$0.13 & -2.31$\pm$0.10 &  37 & -2.31$\pm$0.12 & 13 \\
1234729400256865664 & AE Boo & RRc  & HARPS  & 6630$\pm$75 & 2.04$\pm$0.04 & 2.79$\pm$0.05 & -1.62$\pm$0.14 &  64 & -1.62$\pm$0.10 & 25 \\
3604450388616968576 & AM Vir & RRab & DuPont & 5870$\pm$50 & 1.52$\pm$0.06 & 3.32$\pm$0.06 & -1.70$\pm$0.13 &  87 & -1.70$\pm$0.17 & 25 \\
3626569264033312896 & AS Vir & RRab & DuPont & 6030$\pm$75 & 1.66$\pm$0.10 & 3.44$\pm$0.11 & -1.80$\pm$0.14 &  38 & -1.80$\pm$0.22 & 12 \\
\enddata
\tablecomments{Table \ref{tab:individual_atmospheres} is published in its entirety in the machine-readable format.
A portion is shown here for guidance regarding its form and content.}
\end{deluxetable*}

The final metallicity value for each star was taken as the simple mean of all its measurements. When more than one measurement
was available for a given star, the uncertainty was taken to be the stardard deviation of these measurements. This 
allowed us to compute a typical [Fe/H] uncertainty for each spectrograph by taking the median of these values for each 
instrument. This spectrograph-based median uncertainty was then adopted for the stars with a single spectrum, indicated
by an asterisk on the last column of Tab.~\ref{tab:metallicities}.

The median standard deviations for the uncertainties of the atmospheric parameters are
$\sigma$(T$_{\text{eff}}) =$ 150 $\pm$ 136 K, 
$\sigma$(log(g))$ =$ 0.43 $\pm$ 0.16 dex, and
$\sigma( \xi_{\text{t}}) =$ 0.14 $\pm$ 0.25 km s$^{-1}$. To compute the impact of these values in the metallicity estimates, we have
applied them to V~Ind. We changed the temperature by 150 K, while keeping the other parameters in their final adopted value, 
and registered the difference $\Delta$[Fe/H]$_{T_{\text{eff}}}$ for FeI and FeII lines this change created. We did the same for 
log(g) and $\xi_{\text{t}}$, changing one parameter at a time and keeping the other parameters in their final value. Finally, we 
added in quadrature the three values of $\Delta$[Fe/H] for FeI and and three for FeII. This resulted in a difference of 
0.11 for FeI and 0.14 for FeII when compared to the final adopted atmosphere. Note that we show these values for reference only.
The reason is that the sum in quadrature of correlated errors overestimates the final error, while it also does not take 
systematics into account. The metallicity errors derived by the standard deviation from multiple measurements of the same star 
are, therefore, a more robust uncertainty indicator and that is the value we have adopted, as described in the previous paragraph. 

\begin{figure}
\includegraphics[width=\columnwidth]{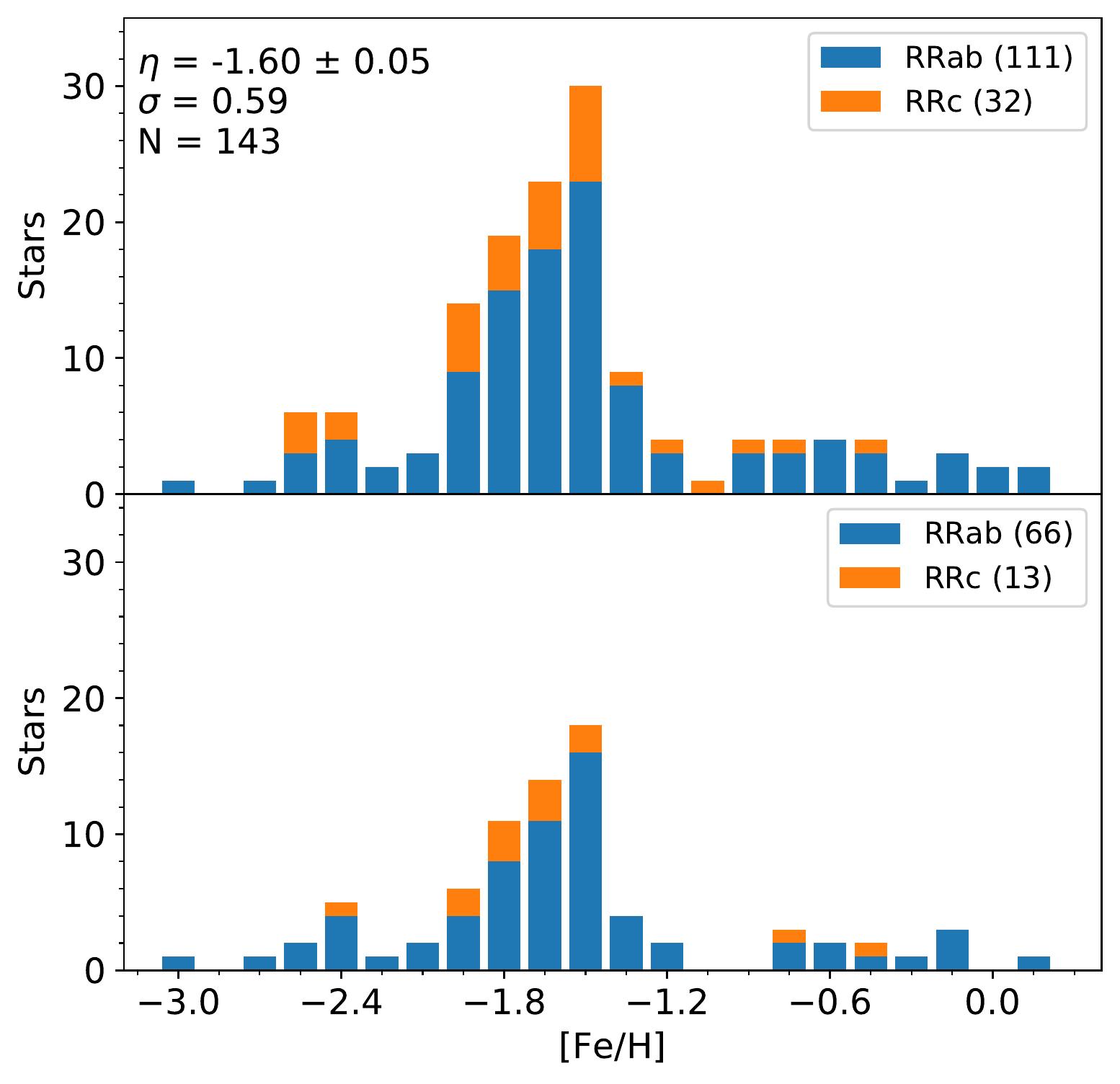}
\caption{\emph{Top:} Metallicity distribution of the 143 calibrating RRLs (RRab, blue; RRc, orange) based on high resolution
spectra. These include new homogeneous measurements for 32 stars with previous measurements in the literature, 
and the results provided by For+Chadid+Sneden
for another 32 stars. \emph{Bottom:} Same as the top, but considering only the 79 stars for which we have provided high resolution
metallicities for the first time. 
\label{fig:calibrators_metallicity_distribution}}
\end{figure}

High resolution metallicities for an additional 32 stars (20 RRab, 12 RRc) were collected from three previous works 
using part of the Du Pont subsample (For+Chadid+Sneden). All their 
uncertanties are the standard deviation derived from multiple metallicity estimates in the same work. Stars with fewer than
two measurements were not considered. The final adopted 
mean metallicities and their uncertainties are listed in Tab.~\ref{tab:metallicities}. The estimates from the present 
work and of these three sources can be treated as a single homogeneous sample and produce a metallicity distribution ranging 
from -3.0 to 0.2 (Fig.~\ref{fig:calibrators_metallicity_distribution}). The entire metallicity range of RRL stars is, thus, covered in our calibrating 
sample of 143 RRL (111 RRab, 32 RRc). 

In order to validate our high resolution measurements, we have compared them to all other such measurements for the same stars
in the literature. This comparison is shown in Fig.~\ref{fig:hires_metallicity_comparison} with the references
listed in Tab.~\ref{tab:compare_metallicities_literature}.

\subsection{High resolution sample}
In addition to the sample described above, we derived metallicities for a mixed mode pulsator from a 
total of ten X-shooter and two FEROS spectra individually, i.e. without stacking. It was not included in the CHR sample
because we aimed to verify whether a mixed-mode pulsator would respond coherently to a $\Delta$S calibration made from a sample
of both RRab and RRc. 

\begin{figure}
\includegraphics[width=\columnwidth]{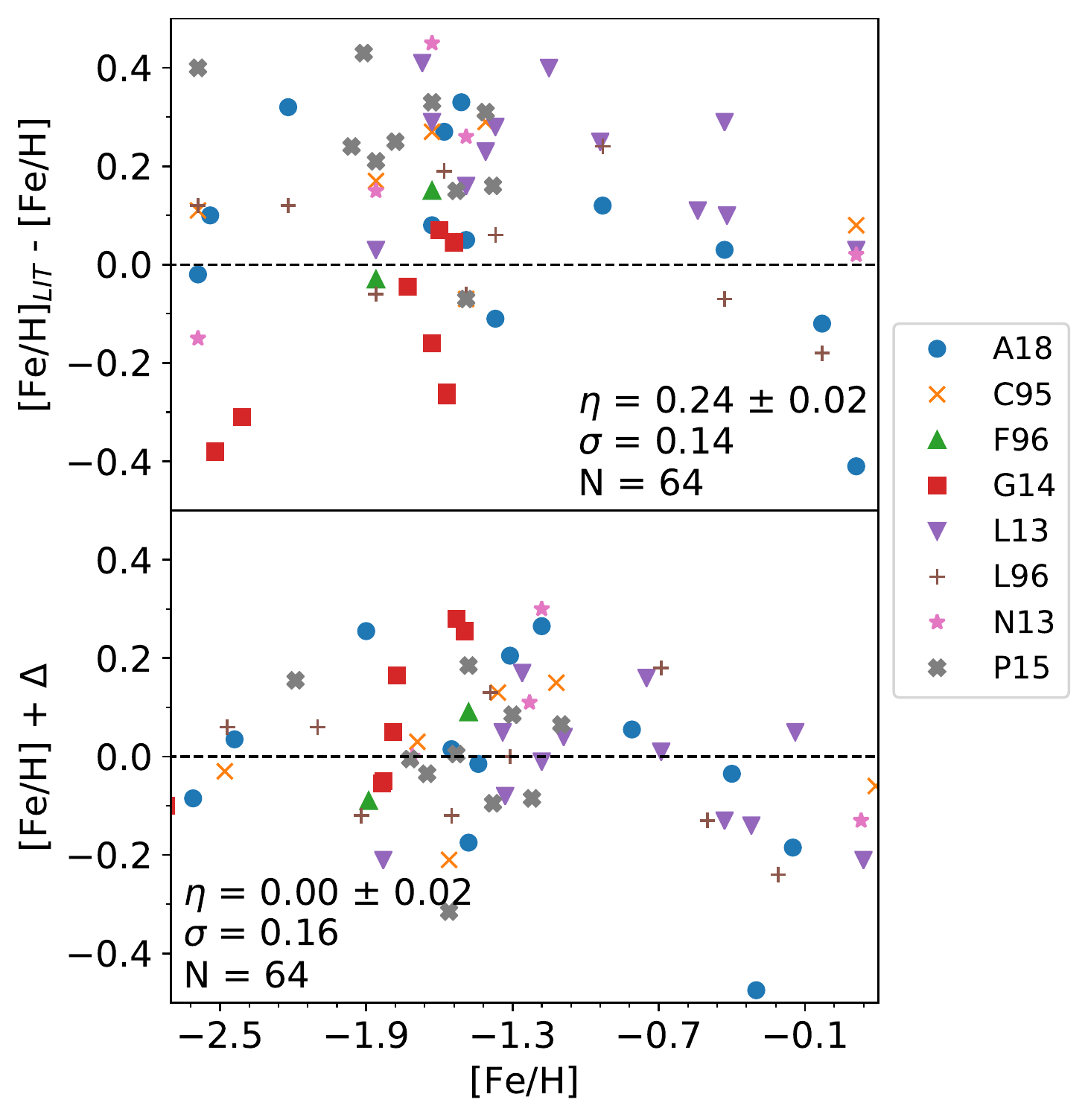}
\caption{\emph{Top:} Comparison of the high resolution spectroscopic metallicities adopted in this work with 
those available in the literature, transformed to the \citet{2009ARAandA..47..481A} Solar scale. \emph{Bottom:} 
The same comparison once the shift $\Delta$ is applied to the literature values, bringing them to our scale. 
The references and shifts are listed in Tab.~\ref{tab:compare_metallicities_literature}.
\label{fig:hires_metallicity_comparison}}
\end{figure}

A total of 134 field RRL have HR iron abundance estimates in the literature. Among these, 64 stars in eight studies
are in common with our CHR sample, while two were in a study without any overlap with this work and were neglected. 
This allowed us to compute the shifts that need to be added to eight studies in
order to bring their metallicity results into our scale (Fig.~\ref{fig:hires_metallicity_comparison}, 
Tab.~\ref{tab:compare_metallicities_literature}). We have applied these shifts to another 67 stars that were in these studies, 
for which we have no metallicity estimate of our own. With the shifts applied, these 67 stars, alongside the CHR sample 
and the RRd mentioned above, form the HR sample
with a total of 211 stars (170 RRab, 40 RRc, 1 RRd). It is the largest, most homogenous sample of high resolution metallicity
 measurements of field RRL stars. Indeed, it is almost a factor of two larger than any previous datasets of the same type in the 
literature \citep{2018ApJ...864...57M,2019ApJ...881..104M}.

\begin{deluxetable}{lcrcrcr} 
\tablecaption{Comparison between high resolution spectroscopic metallicities derived in this work and those 
derived in the literature. The number of stars in common is denoted by N, the median residuals by $\Delta$, 
and their standard deviation by $\sigma$. To bring metallicity values from these works into our scale, the
shift $\Delta$ must be added to them after they are put in the \citet{2009ARAandA..47..481A} Solar scale.
\label{tab:compare_metallicities_literature}}
\tablewidth{\columnwidth}
\tablehead{
\colhead{Source} \hspace{1.2cm} &  \colhead{}\hspace{0.5cm} & \colhead{N} &  \colhead{}\hspace{0.5cm} & \colhead{$\Delta$} &
\colhead{}\hspace{0.5cm}& \colhead{$\sigma$} \\
}
\startdata
A18 & & 12 & & -0.06 & & 0.20 \\
C95 & &  6 & & -0.14 & & 0.12 \\
F96 & &  2 & & -0.06 & & 0.09 \\
G14 & &  8 & &  0.21 & & 0.16 \\
L13 & & 12 & & -0.24 & & 0.12 \\
L96 & &  9 & & -0.06 & & 0.13 \\
N13 & &  5 & & -0.15 & & 0.20 \\
P15 & & 10 & & -0.24 & & 0.14 \\
\hline
Total & & 64 & & -0.11 & & 0.20 \\
\enddata
\tablecomments{
References: 
\citet[][A18]{2018PASP..130b4201A}, \citet[][C95]{1995AJ....110.2319C}, \citet[][F96]{1996AandA...312..957F}, 
\citet[][G14]{2014ApJ...782...59G}, \citet[][L13]{2013RAA....13.1307L}, \citet[][L96]{1996ApJS..103..183L}, 
\citet[][N13]{2013ApJ...773..181N}, \citet[][P15]{2015MNRAS.447.2404P}.
}
\end{deluxetable}

\begin{deluxetable}{lllrl} 
\tablecaption{Adopted metallicities and uncertainties for each individual star in the CHR sample. \label{tab:metallicities}}
\tablewidth{\columnwidth}
\tablehead{
\colhead{GaiaID} & \colhead{Star} & \colhead{Class} & \colhead{[Fe/H]} & \colhead{Source} \\
\colhead{(DR2)} & \colhead{ } & \colhead{ } & \colhead{(dex)} & \colhead{ }
}
\startdata
4224859720193721856 & AA Aql & RRab & -0.42$\pm$0.11 & 1  \\
2608819623000543744 & AA Aqr & RRab & -2.31$\pm$0.04 & 1* \\
1234729400256865664 & AE Boo & RRc  & -1.62$\pm$0.09 & 1* \\
3604450388616968576 & AM Vir & RRab & -1.77$\pm$0.09 & 1  \\
1191510003353849472 & AN Ser & RRab &  0.05$\pm$0.01 & 4  \\
\enddata
\tablecomments{Table \ref{tab:metallicities} is published in its entirety in the machine-readable format.
A portion is shown here for guidance regarding its form and content.
Metallicity and uncertainty sources: this work (1), \citet[][2]{2011ApJS..197...29F}, 
\citet[][3]{2017ApJ...848...68S}, \citet[][4]{2017ApJ...835..187C}. An asterisk denotes that the value was derived from a 
single measurement, and the uncertainty from the typical instrumental uncertainty. See text for details.
}
\end{deluxetable}

\subsection{Low resolution sample} \label{ch:sample_lowres}
Together with the spectra used for the CHR sample we have had at our 
disposal another 298 high resolution spectra of 121 stars (62 RRab, 59 RRc). They were 
not included among the calibrating RRLs because they had low SNR and were 
collected at different phases, making spectra stacking inviable. We degraded and
rebinned their spectra to a spectral resolution R $\approx$ 2,000 and sampling 
$\Delta$log($\lambda$) = 0,0001, in order to mimic the native resolution of the SEGUE-SDSS
spectra. This increased their SNR by a factor of
four to five, approximately. Each star has its own individual [Fe/H]$_{\Delta S}$
computed in the same way the values for the CHR sample were computed
(Sect.~\ref{ch:ds_feh_calibration}). None of these stars has high resolution metallicity 
measurements in the literature.

The main reason why we developed a new $\Delta$S calibration is to apply it to 
low resolution spectra. Thus, we have joined these degraded spectra to the
SEGUE-SDSS low resolution spectra described in Sect.~\ref{ch:lr_dataset} for stars 
that were not added to the HR sample. This provided us with a total of 6,451 
spectra for 5,001 stars (3,439 RRab, 1,562 RRc) that form the low resolution (LR) 
sample.


\section{The new $\Delta $S definition} \label{ch:deltas_definition_measurements}
For the $\Delta $S measurements, all high resolution spectra for the stars in the CHR sample were degraded to 
R $\approx$ 2,000 and rebinned with $\Delta$log($\lambda$) = 0,0001. This ensures that they are similar to the native
low resolution spectra for which the $\Delta$S method was developed. The equivalent widths of the 
lines of interest (Ca II K, H$_\delta$, H$_\gamma$, H$_\beta$) were measured using an updated version of the code 
{\textsc{EWIMH}} \citep[Layden94,][]{2019ApJ...882..169F}. We have adjusted the wavelength limits of the 
$\Delta$S definition in order to increase sensitivity to metallicity.

Absorption lines are formed across a small range of depths in the stellar atmosphere where the thermodynamical 
quantities allow the transition in question to occur. Strong lines may have significant wings that were formed at
different layers and trace different environments. This is especially true in non-metallic, i.e. hydrogen, lines, often causing their 
very wide wings to swallow up metallic lines in their vicinity. The fraction of the equivalent width that comes from extended wings, 
therefore, does not necessarily represent the same thermodynamical quantities nor chemical abundance that the core of
the line does.

The original Layden94 definition considered the cores but not the wings of the hydrogen lines by using a measuring band
with a width of 20 \AA. For the Ca II K line, significantly weaker than the hydrogen lines, it employed a different 
measurement band depending on whether it was determined to be shallow or deep \citep[see Fig.~3 of][]{2019ApJ...882..169F}. 
For the former, the band had a width of 14 \AA, and the latter of 20 \AA. Both wide and narrow definitions 
included the full line with its wings and were careful to avoid nearby lines. 

The Ca II line is flanked by two strong 
absorption features, namely the blend of the Ca II H at 3,968.5 \AA{} with the H$_{\epsilon}$ line at 3,970.0 \AA, and the 
H$_{\zeta}$ line at 3,889.1 \AA. Both of them can have a dramatic effect on the continuum in very hot stars, with wings
reaching as far as within $\approx$15 \AA{} of the central wavelength of the Ca II K line. It is important to note that, as the
hydrogen lines become deeper with increasing temperature, the metallic lines become shallower. Thus, it is reasonable to employ 
different measuring bands in each scenario. In very hot phases, however, the hydrogen lines may be so deep that the 
continuum near the Ca II K line is pulled down, so to say, by their wings. In such cases, even very narrow continuum
bands will still be biased to lower values. For this reason, the original code used short continuum bands and employed 
a safeguard against the lowering of the continuum in hot stars: it allowed the bands to move in search of the maximum 
mean intensity, while keeping their length fixed. Departing from this definition, we tested combinations of the following 
scenarios:

\begin{itemize}
\item Using the original wavelength range for the hydrogen lines, which includes only their cores.
\item Increasing the aforementioned range in order to include their wings.
\item Considering multiple ranges for the Ca II K line, from the innermost 2 \AA{} of the core, to the full line with 
its wings. 
\item Enabling and disabling the change between "wide" and "narrow" Ca II K lines.
\item Changing the continuum band ranges in steps from -20\% to +20\%.
\item Enabling and disabling the movement of the continuum band for the Ca II K line.
\end{itemize}

\begin{figure}
\centering
\includegraphics[width=\columnwidth]{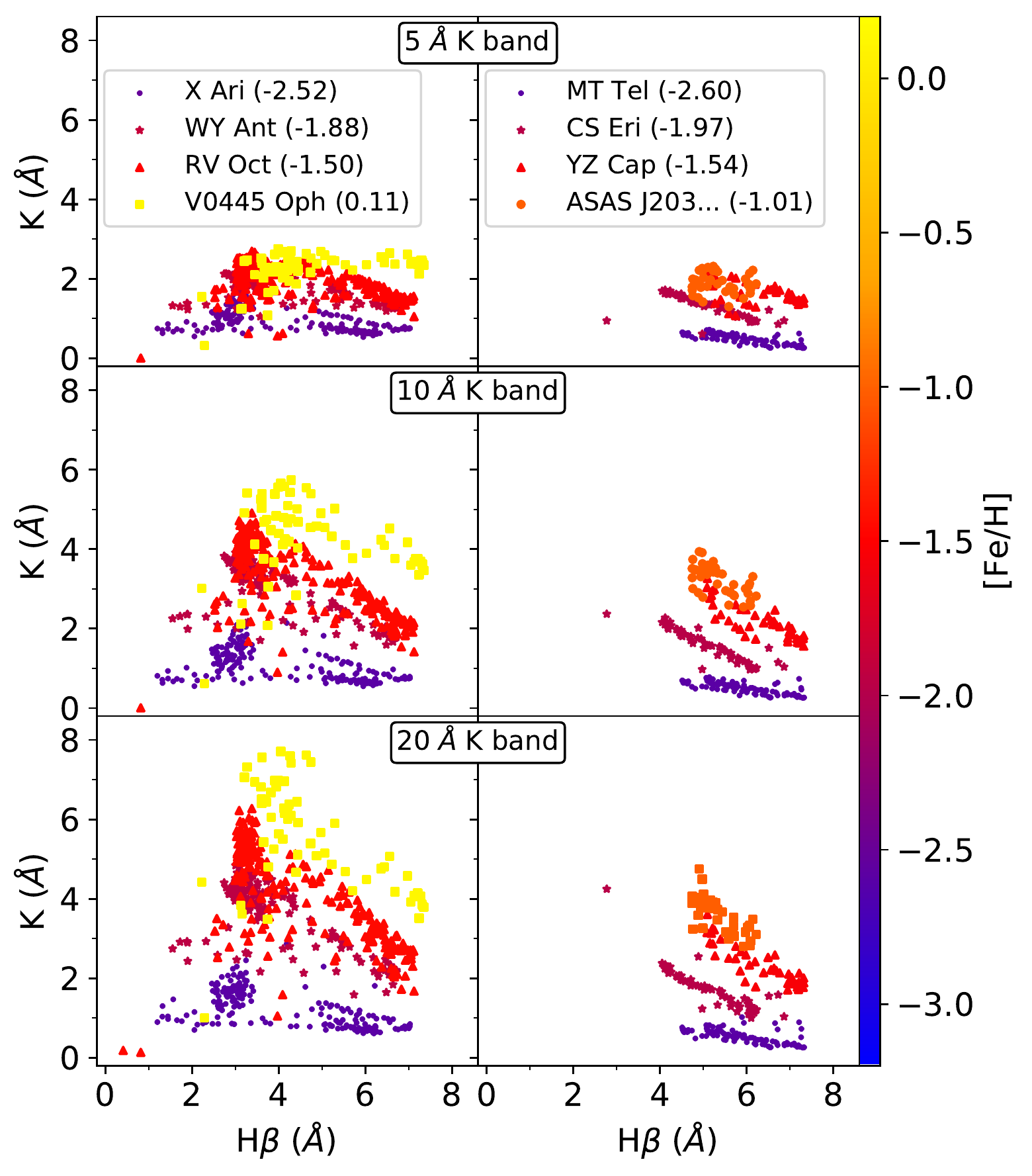}
\caption{The HK plane for H$\beta$ considering different widths (5, 10, and 20 \AA) for the measurement band of the equivalent 
width of the CaII K line. A group of RRab (left) and RRc (right) stars are plotted with markers colored by metallicity 
according to the colorbar at the right edge of the figure.  The considered stars are listed in the legend, with their respective 
high resolution metallicities between parentheses. The name for the RRc ASAS J203145-2158.7 is shortened for convenience.
\label{fig:compare_definitions_H_K_plane}}
\end{figure}

We found that including the wings of the H lines increases scatter without any significant effect on sensitivity 
to metallicity. This is partly due to the presence of metallic lines in the vicinity of the H lines, as noted by Layden94. 
On the other hand, increasing the range of the Ca II K line to always include both core and wings with the full 20 \AA{} 
width for all measurements provided the best 
sensitivity without significant increase in scatter (Fig.~\ref{fig:compare_definitions_H_K_plane}) even for the "narrow" 
lines where continuum noise may creep into the measurements. This may be due to the fact that such noise in our data is 
much lower than in the data used by Layden94. As the change between "narrow" and "wide" Ca II K definitions did not 
cause significant improvement, we have disabled it.

\begin{figure*}
\includegraphics[width=1\textwidth]{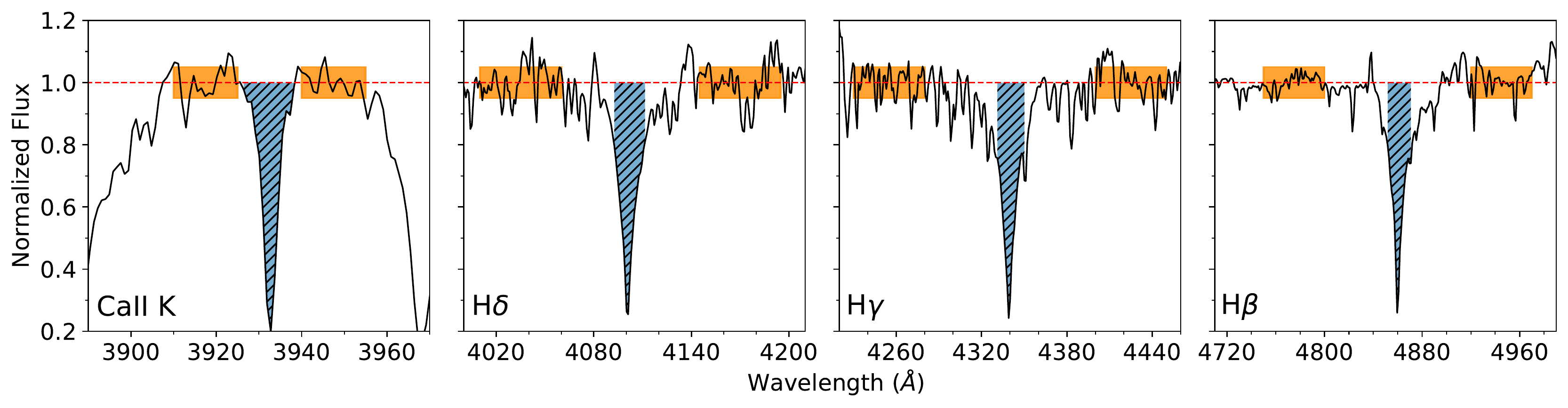}
\caption{Wavelength ranges considered for the $\Delta $S measurements. In black, a du Pont spectrum for AN Ser 
([Fe/H] = 0.05) downgraded to a resolution R = 2,000. The red dashed line indicates the continuum level. The 
hatched blue area denotes the area considered in the equivalent width measurement for each line. The orange shaded 
area indicates the region considered for the continuum level definition. The precise wavelength intervals are
displayed in Tab.~\ref{tab:dS_intervals}.
\label{fig:dS_wavelength_ranges}}
\end{figure*}

Furthermore, we found that the continuum band range introduces only insignificant changes. The only exception is the 
continuum region around the Ca II K line due to the presence of the strong nearby absorption features. Therefore, we 
kept the original 15 \AA{} range that gave enough points to define the continuum even in noisier spectra and saw no need 
to use different band widths on each side of the feature. We did not find a significant improvement when the continuum
band was allowed to move, and so we kept its limits fixed. The final wavelength limits are shown in 
Fig.~\ref{fig:dS_wavelength_ranges} and listed in Tab.~\ref{tab:dS_intervals}. All measurements for the CHR sample
are included in the supplementary Tab.~\ref{tab:deltas_ews}.

\begin{deluxetable}{llcc} 
\tablecaption{Wavelength intervals for the $\Delta $S measurements in the new calibration. 
\label{tab:dS_intervals}}
\tablewidth{\textwidth}
\tablehead{
\colhead{Line} & \colhead{Interval} & \colhead{Start} & \colhead{End} \\ 
\colhead{ } & \colhead{ } & \colhead{(\AA{})} & \colhead{(\AA{})}
}
\startdata
CaIIK & Left continuum   & 3910.00 & 3925.00  \\ 
      & Equivalent width & 3923.67 & 3943.67  \\ 
      & Right continuum  & 3940.00 & 3955.00  \\ 
\hline
H$_{\delta}$    & Left continuum   & 4010.00 & 4060.00  \\ 
      & Equivalent width & 4091.74 & 4111.74  \\ 
      & Right continuum  & 4145.00 & 4195.00  \\ 
\hline
H$_{\gamma}$    & Left continuum   & 4230.00 & 4280.00  \\ 
      & Equivalent width & 4330.47 & 4350.47  \\ 
      & Right continuum  & 4400.00 & 4450.00  \\ 
\hline
H$_{\beta}$    & Left continuum   & 4750.00 & 4800.00  \\ 
      & Equivalent width & 4851.33 & 4871.33  \\ 
      & Right continuum  & 4920.00 & 4970.00  \\ 
\enddata
\end{deluxetable}

\subsection{Changes in the {\textsc{EWIMH}} code}
The EW measurements for the $\Delta$S method can and have been done with multiple codes, among which is the IDL code
{\textsc{EWIMH}}. Alongside the wavelength limits discussed in the previous section, it is crucial to establish that, in 
our calibration, the EW is defined as the area between the observed spectrum and the continuum, derived numerically, 
and not the area under a Gaussian function fit to the data. 

As mentioned above, we have disabled in the \textsc{EWIMH} code the change between "narrow" and "wide" definitions of the Ca II K line, and the 
freedom of movement of the continuum bands around it. These two features were part of the code itself, and should be
disabled in order to comply with our new empirical calibration. The other changes we made to {\textsc{EWIMH}} were done 
in order to make the best use of all available spectra. These changes are not essential to the calibration, although 
we recommend them to be added to whatever code is employed in the EW measurements. 

We adapted the computation of the continuum level and equivalent widths to be more robust against emission 
lines caused by cosmic rays or any instrumental defect that can mimic them such as dead pixels. Equivalent widths are 
defined as the width in wavelength of a rectangle with the height of the continuum that has the 
same area as the area between the absorption feature and the continuum level. Generally speaking, measurement codes 
consider the area of emission features as negative. This means that, when a cosmic ray produces a strong enough 
emission feature, there is a twofold issue: 1) the real area of the line is decreased by the presence of the emission; 
2) if the emission defines an area above the continuum level, this area is subtracted from the equivalent width 
measurement. 

We have introduced a change in the code so that it will discard points with a normalized flux greater than one. This 
addresses the second issue by preventing negative area values and, in many cases, the first as 
well because most emissions do cross the continuum level. The missing value becomes instead a simple interpolation
between its closest neighbors. 

The continuum level is determined by using the mean flux inside the continuum bands at each side of the line of
interest. We have inserted a condition that discards normalized fluxes greater than 1.25 in these bands, thus preventing
emission features or defective pixels from biasing the continuum level determination.


\section{The new $\Delta $S-[Fe/H] calibration} \label{ch:ds_feh_calibration}
A total of 6,327 spectra for 143 stars (111 RRab, 32 RRc) were included in the CHR sample. These are the 
stars for which we have homogeneous HR metallicity measurements (Sect.~\ref{ch:calibrating_sample}). 
They cover a wide range in period, pulsational amplitude (Fig.~\ref{fig:P_Vamp_V_distribution}), and spectroscopic 
metallicity (Fig.~\ref{fig:calibrators_metallicity_distribution}).

We have verified that valid equivalent widths for individual lines in individual spectra remained in the range of
0.01 to 10.00 \AA. Values outside these limits were most often from a distorted line or continuum, and so they were
discarded. To compute the 
$\Delta $S index for each star, we took the median equivalent width of each feature of interest, considering 
all individual spectra for that star. Then, we performed a non-linear least squares fit using the
{\textsc{IDL}} function {\textsc{CURVEFIT}} for a variety of analytical equations in the $\Delta$S index versus 
spectroscopic metallicity [Fe/H] plane. The process consists in applying the equation with free coefficients to
the equivalent widths of interest, deriving one $\Delta$S value for each individual star. These values were then
equaled to their respective high resolution [Fe/H] measurements. The code attempts to minimize the residuals by repeating
this process iteratively as it changes the coefficients.

This is a significant difference between our approach and that of Layden94. Indeed,
Layden94 started from the K versus H plane, where K is the equivalent width of the Ca II K line, and H the average 
EW of the hydrogen lines of interest. While in the K versus H plane there are a series of roughly linear sequences 
with slopes that are metallicity-dependent, we have verified that the resulting $\Delta$S index from a polynomial 
fit done in this plane (i.e. the Layden94 equation, but with free coefficients) does not result in a tighter $\Delta$S 
versus [Fe/H] relation. This is discussed in more detail in Appendix~\ref{ch:appendix_ew_feh_phase}. 

Other than the original Layden94 equation, with and without free coefficients, we have also attempted a variety of 
functions, such as variations of it using the logarithms, polynomials, sums of Gaussians, and Moffat
functions. The minimum scatter was found with a polynomial with the form

\begin{equation} \label{eq:calibration}
\text{[Fe/H]}_{\Delta S} = c_0 + c_1 \text{K} + c_2 \text{H}_{\delta} + c_3 \text{H}_{\gamma} + c_4 \text{H}_{\beta},
\end{equation}

where K, H$_{\delta}$, H$_{\gamma}$, and H$_{\beta}$ are the equivalent widths in angstroms of the Ca II K and H
lines as measured in the updated {\textsc{EWIMH}} code and using the newly defined wavelength ranges 
(Tab.~\ref{tab:dS_intervals}). Here, the $\Delta$S index is already the metallicity estimate [Fe/H]$_{\Delta S}$
and no further transformations are necessary. 

\begin{figure*}
\centering
\includegraphics[width=0.75\textwidth]{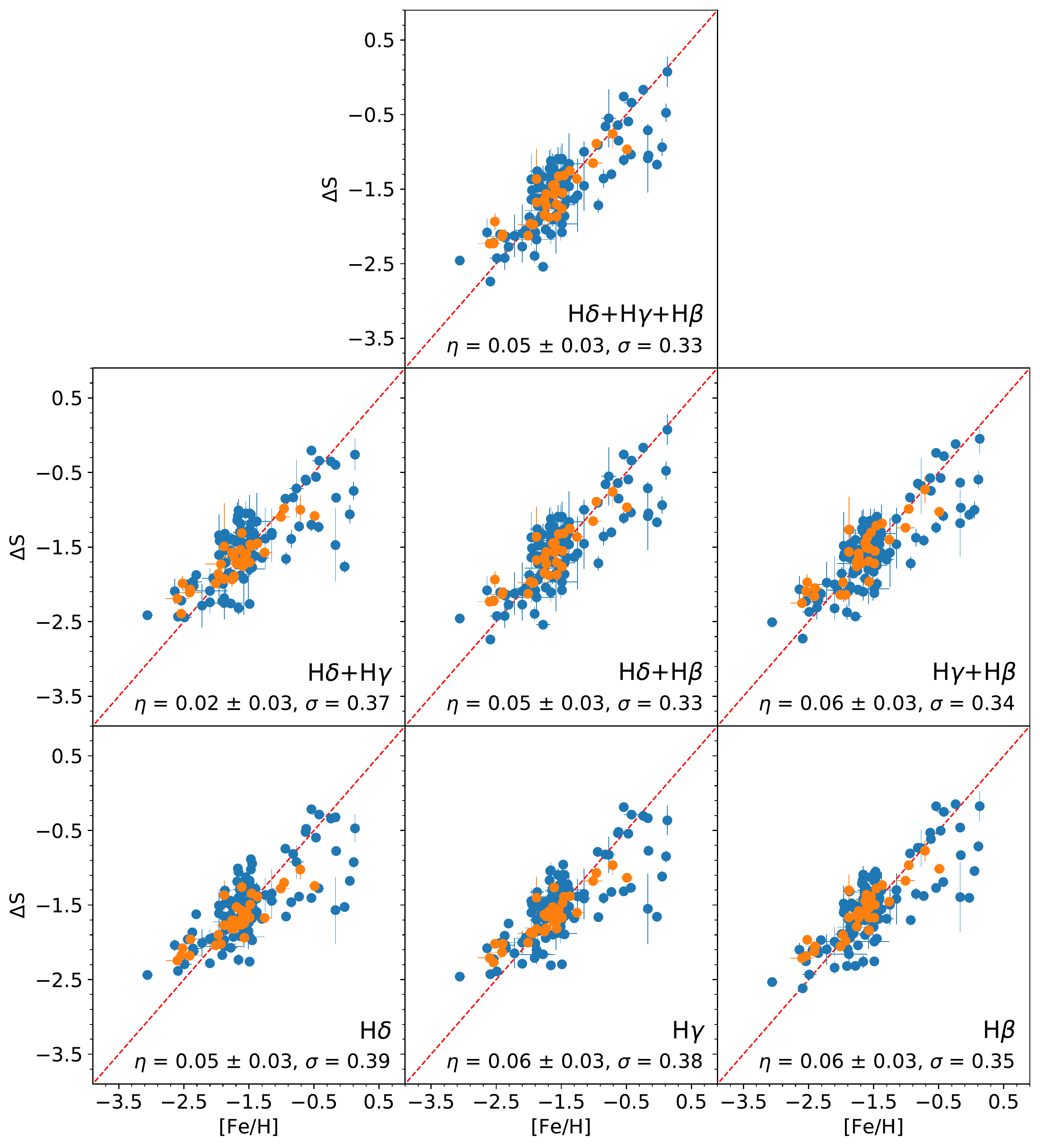}
\caption{The $\Delta$S versus spectroscopic metallicity plane for all combinations of H lines, as indicated in the
annotations inside each plot. The median residuals $\eta$ and the standard deviaton of residuals $\sigma$ 
are also indicated for each case. RRab stars are plotted in blue and RRc in orange. The dashed red lines denote
the identity line.
\label{fig:dS_FeH_finalfit}}
\end{figure*}

The quality of the fit is similar for all combinations of H lines, 
therefore we provide the $c_n$ coefficients for all seven cases in Tab.~\ref{tab:dS_fit_coefficents}. Thus, if 
a given spectrum has only one or two hydrogen lines, [Fe/H]$_{\Delta S}$ can still be estimated using the appropriate
coefficients, with the missing H line having its corresponding $c_n$ coefficient set to zero. The $\Delta$S versus 
spectroscopic metallicity plane for all combinations is shown in Fig.~\ref{fig:dS_FeH_finalfit}, while the corresponding
values are listed in Tab.~\ref{tab:dS_measurements}. Throughout this 
work, all comparisons performed against our own HR estimates or literature values were performed using the
[Fe/H]$_{\Delta S}$ equation for the full set of hydrogen lines.

\begin{deluxetable*}{lrrrrrr} 
\tablecaption{Coefficients for the new $\Delta$S equation (Eq.~\ref{eq:calibration}) for all combinations of hydrogen lines and the 
corresponding standard deviation of the residuals $\sigma$.
\label{tab:dS_fit_coefficents}}
\tablewidth{\textwidth}
\tablehead{
\colhead{H lines} & \colhead{c$_0$} & \colhead{c$_1$} & \colhead{c$_2$} & \colhead{c$_3$} & \colhead{c$_4$}  & \colhead{$\sigma$} \\
}
\startdata
H$\delta$, H$\gamma$, H$\beta$ & -3.84323 $\pm$ 0.02438 & 0.36828 $\pm$ 0.08481 & -0.22182 $\pm$ 0.11325 &  0.00433 $\pm$ 0.08793 & 0.51481 $\pm$ 0.18314 & 0.33 \\
H$\delta$, H$\gamma$           & -3.75381 $\pm$ 0.02682 & 0.39014 $\pm$ 0.09433 & -0.19997 $\pm$ 0.10267 &  0.38916 $\pm$ 0.20318 & \multicolumn{1}{c}{--} & 0.37 \\
H$\delta$, H$\beta$            & -3.84160 $\pm$ 0.02302 & 0.36798 $\pm$ 0.05519 & -0.21936 $\pm$ 0.07134 & \multicolumn{1}{c}{--} & 0.51676 $\pm$ 0.17720 & 0.33 \\
H$\gamma$, H$\beta$            & -3.79074 $\pm$ 0.02462 & 0.35889 $\pm$ 0.07550 & \multicolumn{1}{c}{--} & -0.21997 $\pm$ 0.08967 & 0.50469 $\pm$ 0.18582 & 0.34 \\
H$\delta$                      & -3.48130 $\pm$ 0.02690 & 0.36105 $\pm$ 0.02689 &  0.14403 $\pm$ 0.19890 & \multicolumn{1}{c}{--} & \multicolumn{1}{c}{--} & 0.38 \\
H$\gamma$                      & -3.70799 $\pm$ 0.02682 & 0.38127 $\pm$ 0.02831 & \multicolumn{1}{c}{--} &  0.17973 $\pm$ 0.20453 & \multicolumn{1}{c}{--} & 0.38 \\
H$\beta$                       & -3.92067 $\pm$ 0.02393 & 0.38194 $\pm$ 0.03126 & \multicolumn{1}{c}{--} & \multicolumn{1}{c}{--} & 0.25898 $\pm$ 0.18516 & 0.35 \\
\enddata
\end{deluxetable*}

\begin{deluxetable*}{llrrrrrrrr} 
\tablecaption{The [Fe/H]$_{\Delta S}$ measurements for the CHR sample. Columns four to ten indicate between square brackets the combinations of hydrogen 
lines considered in the measurement.\label{tab:dS_measurements}}
\tablewidth{\columnwidth}
\tablehead{
\colhead{GaiaID} & \colhead{Class} &  \colhead{N$_{\text{spec}}$} &
\colhead{[H$\delta$, H$\gamma$, H$\beta$]} &  
\colhead{[H$\delta$, H$\gamma$]} & \colhead{[H$\delta$, H$\beta$]} & \colhead{[H$\gamma$, H$\beta$]} & 
\colhead{[H$\delta$]} & \colhead{[H$\gamma$]} & \colhead{[H$\beta$]} \\
\colhead{(DR2)} & \colhead{ } &  \colhead{ } &
\colhead{(dex)} &  
\colhead{(dex)} & \colhead{(dex)} & \colhead{(dex)} & 
\colhead{(dex)} & \colhead{(dex)} & \colhead{(dex)} 
}
\startdata
15489408711727488   & RRab & 194 & -2.74$\pm$0.07 & -2.43$\pm$0.05 & -2.74$\pm$0.07 & -2.73$\pm$0.07 & -2.38$\pm$0.02 & -2.43$\pm$0.02 & -2.62$\pm$0.03 \\
77849374617106176   & RRab & 6   & -1.62$\pm$0.15 & -1.56$\pm$0.14 & -1.61$\pm$0.15 & -1.56$\pm$0.15 & -1.44$\pm$0.07 & -1.51$\pm$0.08 & -1.56$\pm$0.08 \\
80556926295542528   & RRc  & 3   & -1.36$\pm$0.41 & -1.49$\pm$0.58 & -1.36$\pm$0.41 & -1.27$\pm$0.45 & -1.37$\pm$0.20 & -1.40$\pm$0.28 & -1.30$\pm$0.21 \\
630421935431871232  & RRab & 3   & -1.85$\pm$0.51 & -1.93$\pm$0.48 & -1.85$\pm$0.51 & -1.80$\pm$0.47 & -1.84$\pm$0.23 & -1.88$\pm$0.23 & -1.84$\pm$0.25 \\
1167409941124817664 & RRc  & 1   & -1.68$\pm$0.00 & -1.93$\pm$0.00 & -1.68$\pm$0.00 & -1.56$\pm$0.00 & -1.77$\pm$0.00 & -1.82$\pm$0.00 & -1.67$\pm$0.00 \\
\enddata
\tablecomments{Table \ref{tab:dS_measurements} is published in its entirety in the machine-readable format.
A portion is shown here for guidance regarding its form and content.
}
\end{deluxetable*}

The behavior of the Ca II K and hydrogen lines is peculiar in phases between 0.9 and 0.1, in particular in some RRab
stars with larger pulsational amplitudes \citep[e.g.][]{2014A&A...565A..73G,2016A&A...587A.134G}. In previous calibrations 
of the $\Delta$S method, these phases were neglected 
due to their association with the formation and propagation of shock waves, but this effect has never been investigated 
on a quantitative basis for the $\Delta$S method. We have verified that the removal of this phase interval from the
CHR sample results in an improvement of only 0.02 in the scatter. This is also discussed in 
Sect.~\ref{ch:appendix_ew_feh_phase}. Removing the RRc had similarly negligible effects. 

In order to properly classify the pulsation mode of an RRL star, multiple photometric observations and 
possibly a light curve template are needed so that a light curve with enough phase points can be built. Synchronizing 
observation time with pulsation phase requires both this detailed study and the availability of the telescope at the 
specific time window. If a star displays a mixed mode, or any other pulsational irregularity, this synchronization may not be 
possible at all. This is the reason we are providing a single calibration for both RRab and RRc at all phases. The 
negligible effect on the scatter when removing specific phases or separating the RRL by pulsation mode shows that
previous studies of the RRLs of interest, in order to classify them and derive precise light curves, are unnecessary with this 
calibration.

\begin{figure}
\includegraphics[width=\columnwidth]{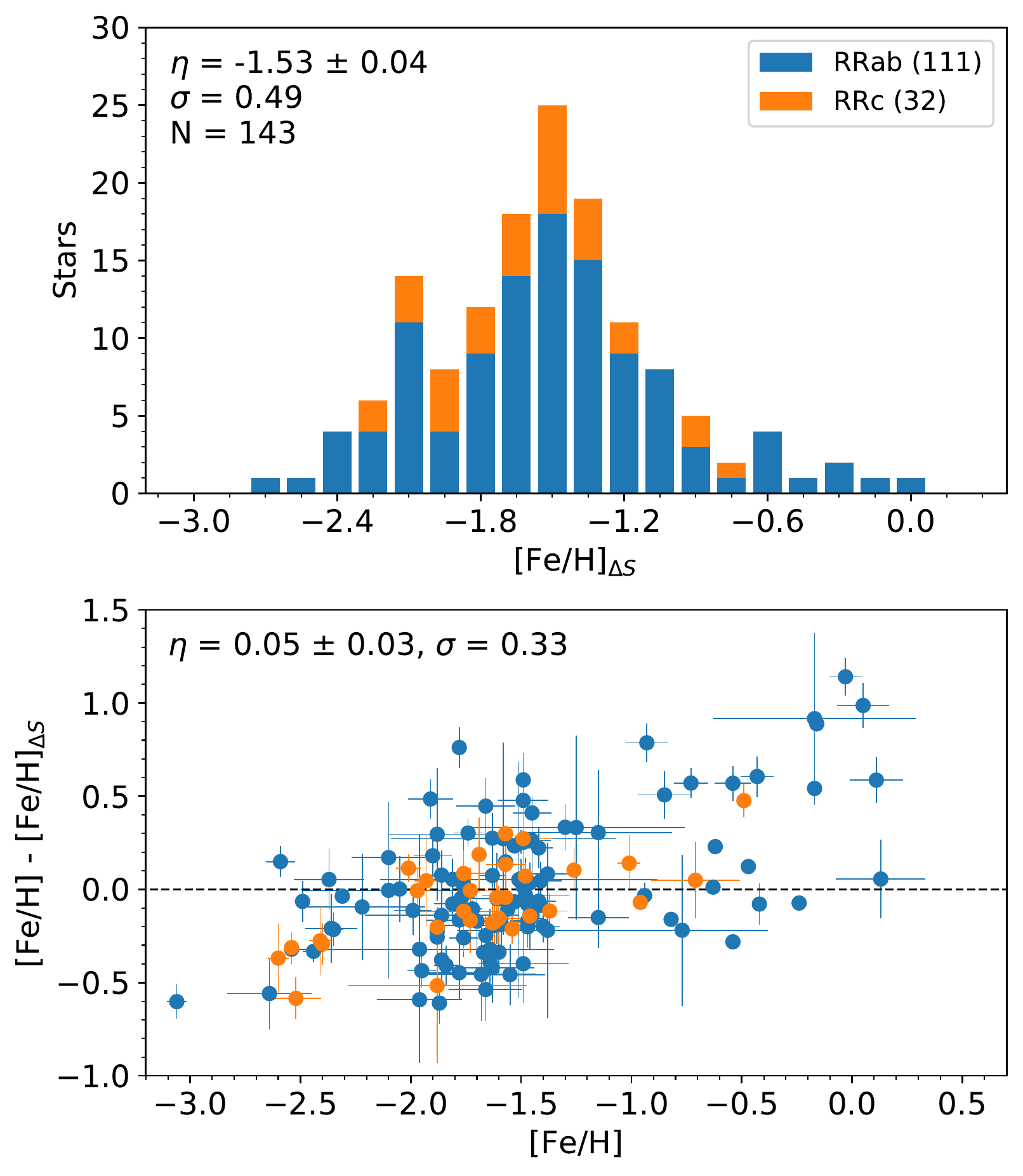}
\caption{\emph{Top:} The [Fe/H]$_{\Delta S}$ distribution for the CHR sample. 
For reference, the distribution of high resolution measurements for the same sample is shown 
in the top panel of Fig.~\ref{fig:calibrators_metallicity_distribution}.
\emph{Bottom:} Residuals of the new $\Delta $S calibration versus spectroscopic metallicity. 
The RRab are plotted in blue, and the RRc in orange. 
\label{fig:dS_metallicity_distribution_calibrators}}
\end{figure}

Finally, the residuals do not show any clear trend with period or amplitudes, but are greater in the 
high metallicity regime. This is evident in the bottom panel of Fig.~\ref{fig:dS_metallicity_distribution_calibrators}, which
shows the difference between the HR [Fe/H] measurement and the low resolution [Fe/H]$_{\Delta S}$ estimate,
plotted against the HR [Fe/H]. The larger residuals for metal rich stars are also evident in Fig.~\ref{fig:dS_FeH_finalfit}, where a
group of about five stars deviates from the main relation, displaying lower [Fe/H]$_{\Delta S}$ values than expected.
More data in the high metallicity regime are required to further constrain the nature of the spread in 
$\Delta$S values.

To constrain the accuracy of the new $\Delta$S calibration we estimated the 
metallicity distribution of the CHR sample and we found a median 
$\eta$ = -1.53$\pm$0.00 and a standard deviation of $\sigma$ = 0.49   
(Fig.~\ref{fig:dS_metallicity_distribution_calibrators}, top panel). This means that the $\Delta$S 
calibration can recover the metallicity distribution based on high 
resolution estimates for the same stars. Indeed, the latter has a median 
[Fe/H] = -1.60$\pm$0.00 and $\sigma$ = 0.59  
(Fig.~\ref{fig:calibrators_metallicity_distribution}). The median residual
between both estimates is $\eta$ = 0.05$\pm$0.03, with $\sigma$ = 0.33.

\section{Validation} \label{ch:validation}

To verify the accuracy of the iron abundances based on the new $\Delta$S 
calibration we performed three independent tests using iron abundances based on
literature values derived from both high and low resolution spectra.

\subsection{Comparison with high resolution estimates in the literature}

\begin{figure}
\includegraphics[width=\columnwidth]{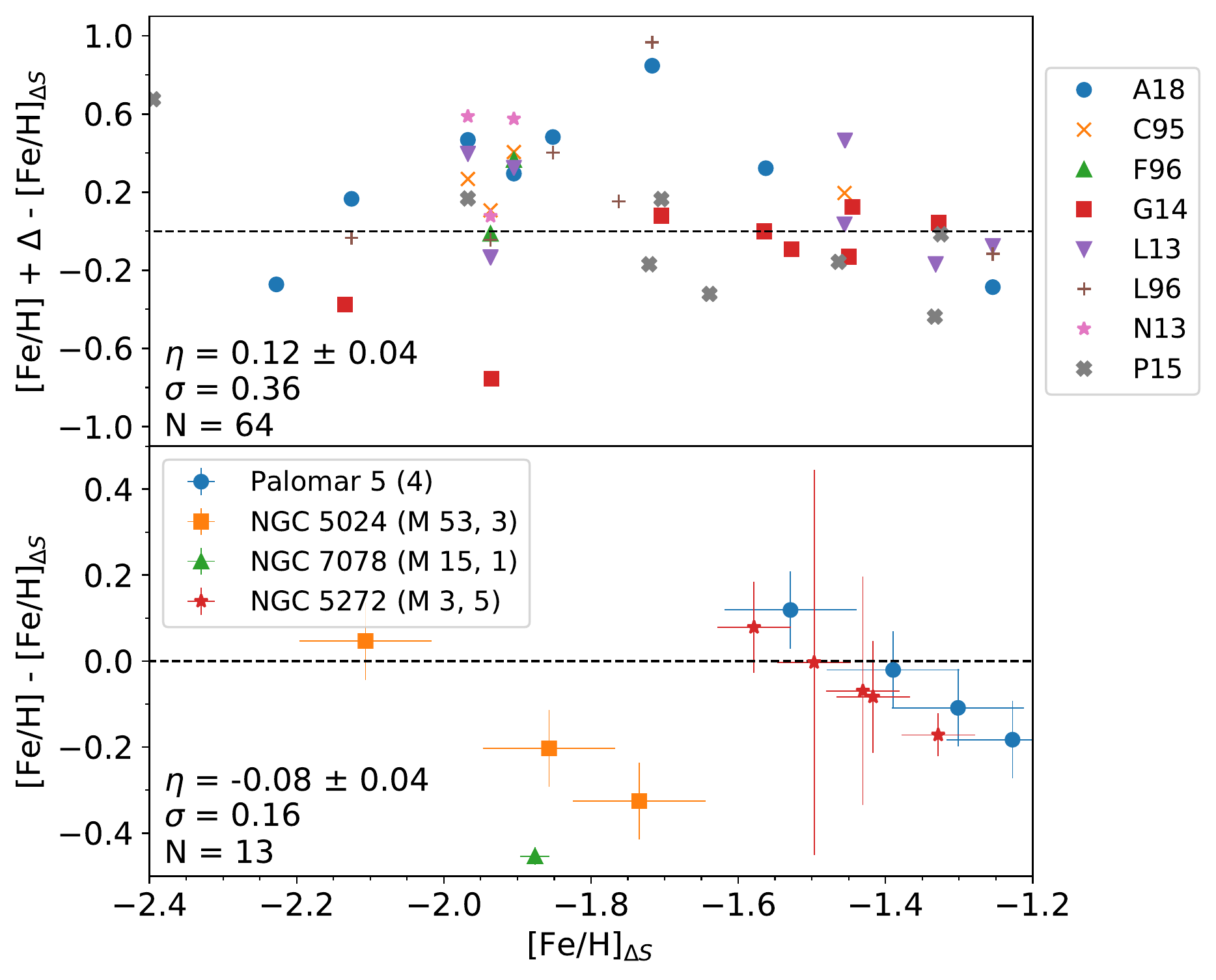}
\caption{Comparison between the new [Fe/H]$_{\Delta S}$ values and literature high resolution metallicities for RRL stars.
\emph{Top:} Field RRL with literature values brought to our scale by the addition of the corresponding shift ($\Delta$).
The references and shifts are listed in 
Tab.~\ref{tab:compare_metallicities_literature}. \emph{Bottom:} Globular cluster RRL considering the cluster metallicities
derived by \citet{2009A&A...508..695C}, which provides their value on a scale based on high resolution spectra. No shift 
has been applied to these values.
\label{fig:dS_lit_hires_compare}}
\end{figure}

Concerning the iron abundances based on high resolution estimates, we performed two tests. 
First, we took advantage of the iron abundances for field RRLs based on HR spectra available 
in the literature. These have already been discussed during the validation of our high resolution 
metallicities (Sect.~\ref{ch:deltas_definition_measurements}), and the shifts ($\Delta$) 
to move them in the current metallicity scale are listed in 
Tab.~\ref{tab:compare_metallicities_literature}. They are included in the HR sample and contain
64 objects in common with the CHR sample. We found that 
the median difference between the new 
[Fe/H]$_{\Delta S}$ and the literature high resolution estimates re-scaled into our scale 
is $\eta$=0.12$\pm$0.04, $\sigma$=0.36 (Fig.~\ref{fig:dS_lit_hires_compare}, top panel). 

Second, we computed $\Delta$S metallicities for 12 RRLs that belong to Galactic globular clusters (GCs). 
The GC metallicities were provided by \citet{2009A&A...508..695C} in a new scale they constructed from high 
resolution estimates. No scale transformation was performed. The residuals of the comparison resulted in a 
median $\eta$ = -0.08 $\pm$ 0.04 and $\sigma$ = 0.16  (Fig.~\ref{fig:dS_lit_hires_compare}, bottom panel).

\subsection{Comparison with low resolution estimates in the literature}

\begin{figure}
\includegraphics[width=\columnwidth]{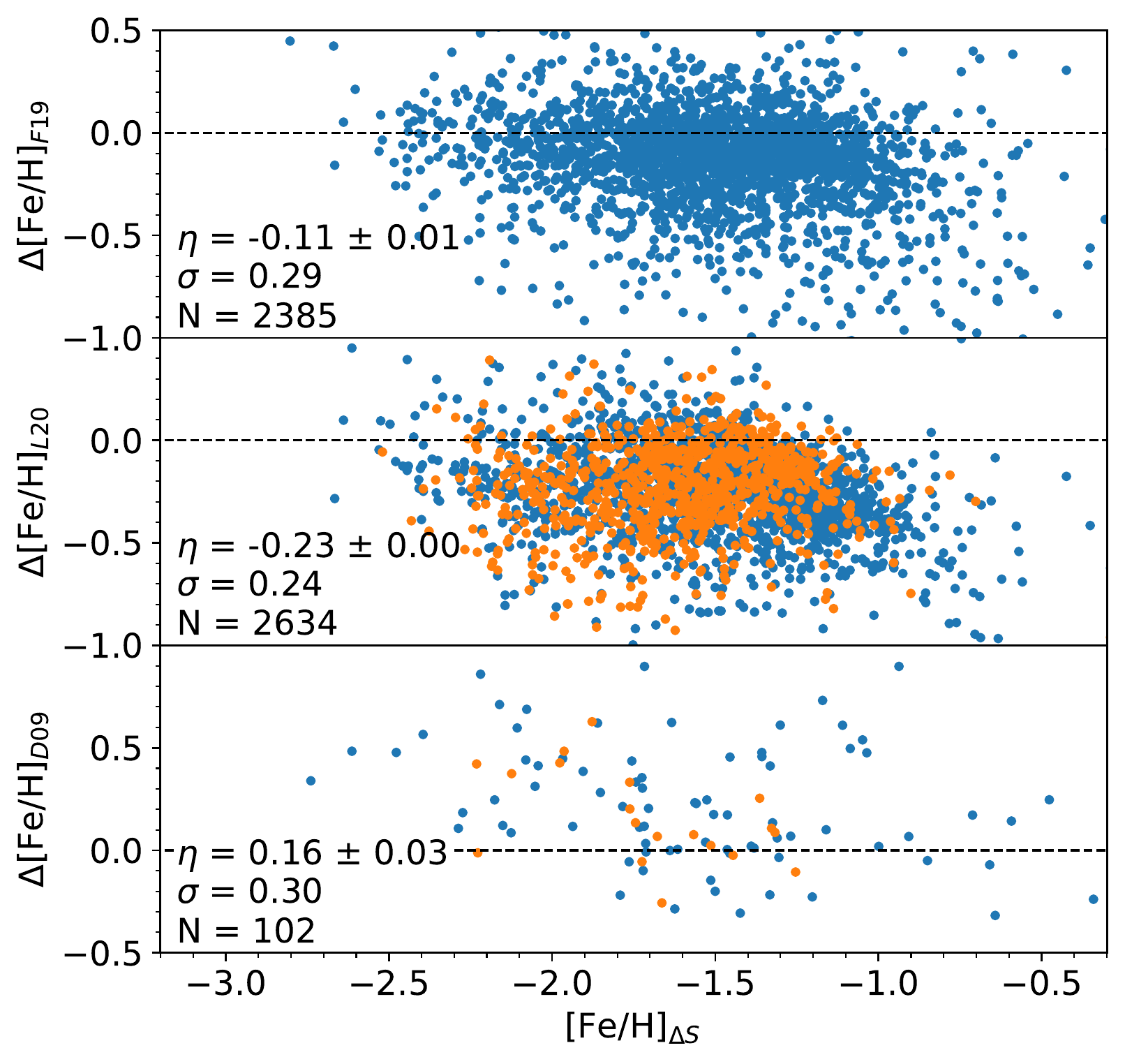}
\caption{Difference $\Delta$[Fe/H] - [Fe/H]$_{\Delta S}$ between the new [Fe/H]$_{\Delta S}$ values and those derived by 
\citet{2019ApJ...882..169F} with the SEGUE-SDSS dataset using the Layden94 definition ($\Delta$[Fe/H]$_{F19}$, top panel), 
\citet{2020ApJS..247...68L} with the SEGUE-SDSS and LAMOST datasets but using low resolution spectral matching ($\Delta$[Fe/H]$_{L20}$, middle panel), 
and the \citet{2009MNRAS.396..553D} compilation ($\Delta$[Fe/H]$_{D09}$, bottom). The RRab are plotted in blue and the RRc in orange.
\label{fig:dS_fabrizio_liu_dambis_compare}}
\end{figure}

The RRab in the SEGUE-SDSS dataset were investigated in detail by F19 using the Layden94 definition. 
Considering the 2,385 RRab in common between both works, their estimates produce a metallicity distribution with median 
[Fe/H]$_{F19}$ = -1.60 $\pm$ 0.01 and $\sigma$ = 0.40. Meanwhile, the new definition applied to the same stars provides 
[Fe/H]$_{\Delta S}$ = -1.48 $\pm$ 0.01 and $\sigma$ = 0.39. The top panel of 
Fig.~\ref{fig:dS_fabrizio_liu_dambis_compare} shows the residuals of this comparison.

\citet{2020ApJS..247...68L} constructed a large catalog of Galactic RRL with metallicities estimated from low resolution
spectral matching. They made use of spectra from both SEGUE-SDSS and the LAMOST Experiment for Galactic Understanding and
Exploration \citep{2012RAA....12..723Z}. A total of 2,634 of their stars are included in our SEGUE-SDSS sample, allowing us 
to make a direct comparison 
between the two different approaches to metallicity estimation. Interestingly, their spectral matching technique relies on 
using precisely the Ca II K line, and the same three Balmer lines of the $\Delta$S method. The median difference between
their values and the ones derived in this work is $\eta$ = -0.23 $\pm$ 0.00 with $\sigma$ = 0.24 
(Fig.~\ref{fig:dS_fabrizio_liu_dambis_compare}, middle panel). The residuals are very similar for both the RRc 
($\eta$ = -0.21 $\pm$ 0.01, $\sigma$ = 0.23) and the RRab ($\eta$ = -0.23 $\pm$ 0.01, $\sigma$ = 0.24). Note that the
high resolution estimates they adopted to validate their method are, for 18 out of 19 stars, \citet{2015MNRAS.447.2404P} and
\citet{2013ApJ...773..181N}. These two works are included in our high resolution comparison and also present a similar shift 
(Fig.~\ref{fig:hires_metallicity_comparison} and Tab.~\ref{tab:compare_metallicities_literature}). Therefore, this scale
difference of $\approx$-0.2 between their results and ours is to be expected.

Finally, no validation would be complete without considering \citet{2009MNRAS.396..553D} (D09), a compilation of RRL metallicities
widely used in the literature. In order to keep the metallicity scale as homogeneous as possible, D09 adopted the values from Layden94
and \citet{1996AJ....112.2110L} wherever they were available, complementing them with other sources after these were transformed into the 
same Layden94 scale, which is the \citet{1984ApJS...55...45Z} scale. In our comparison of 102 stars in common between their work and the 
new [Fe/H]$_{\Delta S}$ estimates
for both the calibrating and LR samples, we found a shift of $\eta$=0.16$\pm$0.03, $\sigma$=0.30. However, to avoid having metallicities
 transformed into one scale, and then transformed once more into another scale, we will provide the corrections for the different D09 
sources separately in a forthcoming paper. 

\subsection{Comparison with other low resolution spectrographs}

We investigated the accuracy of the new [Fe/H]$_{\Delta S}$ calibration in 
estimating the metallicity of individual RRLs when applied to native low resolution 
spectra collected with different instrument and telescope combinations. In particular, 
we have 
secured a set of LAMOST DR6 spectra for RRLs in common either with the LR sample 
or with the HR sample. The new calibration was only applied to LAMOST 
spectra with SNR larger than ten in the region of the CaII K line. Afterwards, 
we visually inspected all the spectra with SNR between ten and twenty for a 
supplementary quality check before including them in the current analysis. 

\begin{figure}
\includegraphics[width=\columnwidth]{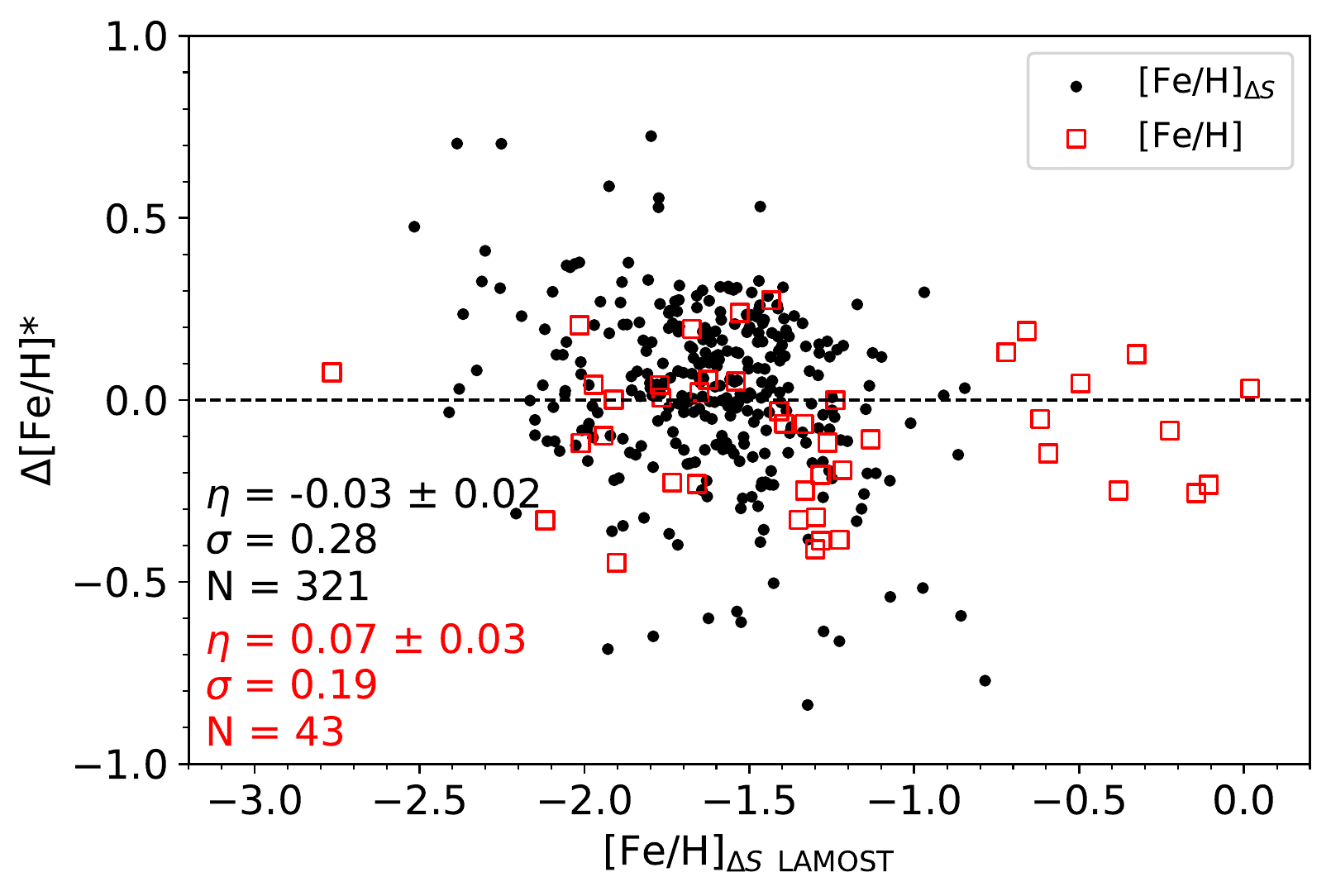}
\caption{Residuals between [Fe/H]$_{\Delta S}$ estimated with the new calibration applied to the RRLs of the LAMOST dataset, 
and the corresponding metallicities of the LR (black dots) and the HR (red squares) samples. The medians $\eta$, standard deviations 
$\sigma$, and sample sizes N are shown in black for the LR results, and in red for the HR results.
\label{fig:dS_FeH_SEGUE_LAMOST_EFOSC_compare}}
\end{figure}

This resulted in a LAMOST sample of 569 low resolution spectra for 364 RRLs  
(203 RRab, 161 RRc). For 321 of these RRLs, we have [Fe/H]$_{\Delta S}$ 
estimates coming from the LR sample, while for 43 of them we have [Fe/H] 
measurements in the HR sample. The [Fe/H]$_{\Delta S}$ values derived 
from LAMOST spectra using the three hydrogen lines show no trend when 
compared to iron abundances based on the HR measurements 
(Fig.~\ref{fig:dS_FeH_SEGUE_LAMOST_EFOSC_compare}). 
The same outcome applies to the comparison with the LR sample, indeed,
the residuals show a larger scatter ($\sigma$=0.28 dex versus
$\sigma$=0.19 dex), but they are within the typical uncertainties of the
method. 

These good correlations indicate that the new [Fe/H]$_{\Delta S}$ 
calibration when
applied to low resolution spectra collected with diﬀerent spectrographs
provides metal abundances that are on the same scale.The same outcome 
applies to the comparison with the LR sample, 
indeed, the residuals show a larger scatter ($\sigma$=0.28 dex versus 
$\sigma$=0.19 dex), but they are within the typical uncertainties of the method.

In passing we also note that the SEGUE-SDSS and the LAMOST datasets when considered 
together, include eleven RRLs with a number of spectra randomly collected ranging 
from five to 15. To further investigate possible variations in metal abundances 
along the pulsation cycle, we applied the new calibration to these low resolution 
spectra collected with different spectrographs. We found a median standard 
deviation per RRL of $\sim$0.15 dex, in very good agreement with results based on 
the degraded high resolution spectra (see Tab.~\ref{tab:dS_measurements}).


\section{Application of the new $\Delta$S calibration} \label{ch:application}
\subsection{Metallicity distribution of field RRL}
Our new empirical [Fe/H]$_{\Delta S}$ calibration allowed us, for the first time in the literature, to probe the metallicity distribution
of the Galactic halo by applying the $\Delta$S method to both RRab and RRc stars. For this end, we employed the LR sample. 
It contains no stars in common with the HR sample, which we will analyse separately.

\begin{figure}
\includegraphics[width=\columnwidth]{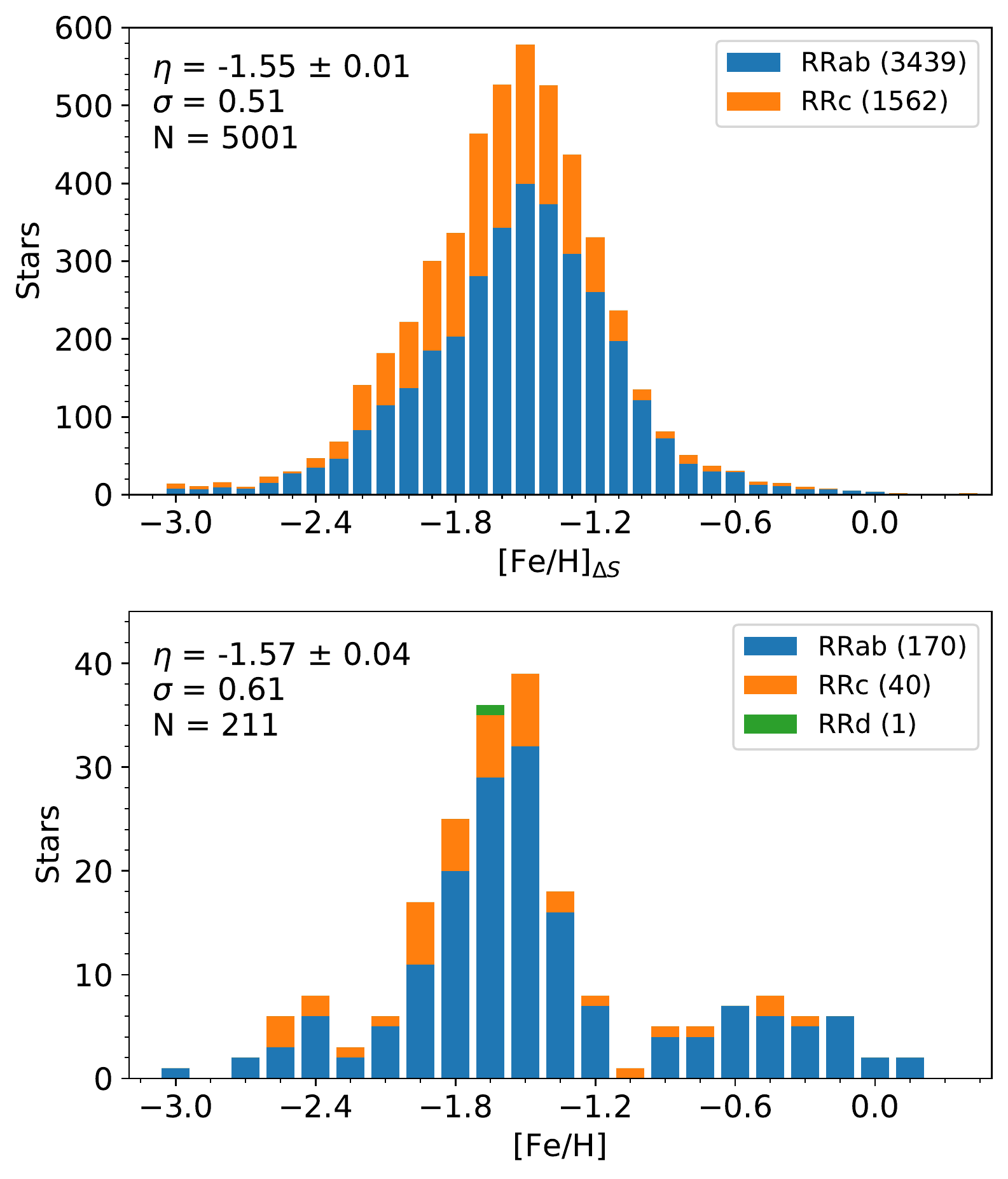}
\caption{\emph{Top:} Distribution of metallicities based on the new $\Delta$S definition for LR sample, consisting of 5,001 
Galatic halo RRL stars (3,439 RRab, 1,562 RRc). \emph{Bottom:} The HR sample, consisting of 211 RRL (170 RRab, 40 RRc, 1 RRd)
from this work and the literature, with the latter brought into our scale.
The RRab, RRc, and RRd are plotted, respectively, in blue, orange, and green.
\label{fig:dSlowres_FeHhires_metallicity_distribution}}
\end{figure}

The resulting [Fe/H]$_{\Delta S}$ distribution has a median $\eta$ = -1.55 $\pm$ 0.01 and $\sigma$ = 0.51 
(Fig.~\ref{fig:dSlowres_FeHhires_metallicity_distribution}, top panel). The RRc display generally lower metallicities. Separating 
fundamental and first overtone pulsators results in medians that differ by 0.12, with the RRab peaking at 
[Fe/H]$_{\Delta S}$ = -1.51 $\pm$ 0.01, with $\sigma$ = 0.50. Meanwhile, the values for the RRc are 
[Fe/H]$_{\Delta S}$ = -1.63 $\pm$ 0.01 and $\sigma$ = 0.50.

The lower metallicity distribution for the RRc is also found in the results of \citet{2020ApJS..247...68L}. 
Considering the stars in common between their work and ours, but their metallicity results, the RRab present $\eta$ = -1.69$\pm$0.01, 
$\sigma$ = 0.35, while for the RRc the values are $\eta$ = -1.80 $\pm$ 0.01, $\sigma$ = 0.38. As mentioned in the previous section, 
there is a zero point shift of about 0.2 between their scale and ours that begins with the high resolution reference scale each 
work adopted. 

\begin{figure}
\includegraphics[width=\columnwidth]{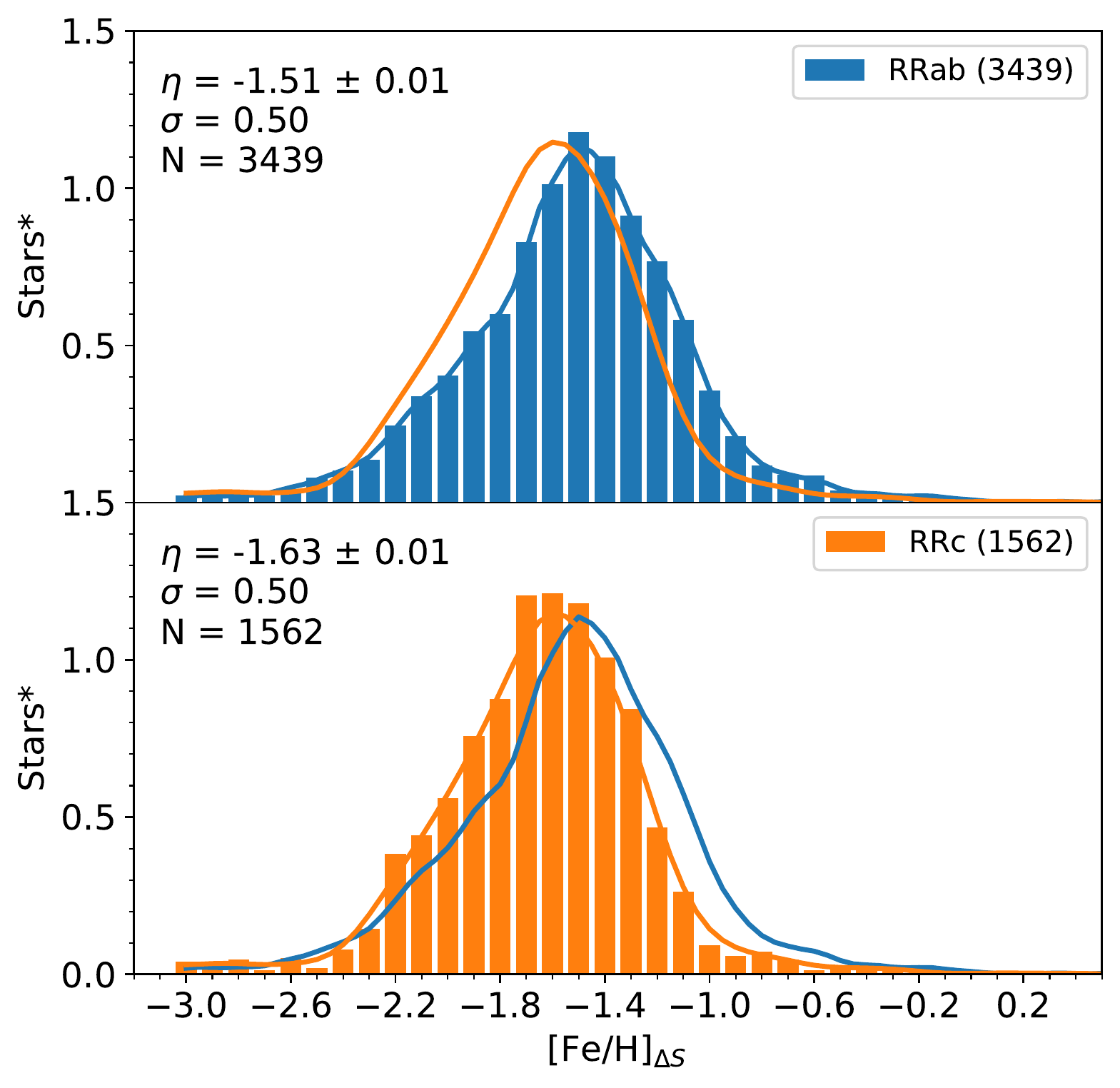}
\caption{Area-normalized histograms of [Fe/H]$_{\Delta S}$ for the LR sample. The RRab are plotted in blue in the top panel, and the RRc 
in orange in the bottom panel. The full lines are the Gaussian-smoothed distributions for the RRab and the RRc, in blue and orange,
respectively. They are shown together on both panels for comparison.
\label{fig:dSlowres_FeHhires_metallicity_distribution_RRab_RRc}}
\end{figure}

Interestingly, the metallicity distribution as a whole is asymmetrical and cannot be adequately described by a single Gaussian, as
is also the case for the kinematic distribution of the field halo stars \citep[e.g.][]{2019MNRAS.486..378L}.
The slope in the metal-rich regime is steeper than in its metal-poor counterpart. In this context it is worth mentioning that the
difference between the metallicity distribution of RRab and RRc is further supported by their asymmetry (skewness). 
Fig.~\ref{fig:dSlowres_FeHhires_metallicity_distribution_RRab_RRc} shows that the metal-rich tail of both RRab and RRc agrees quite 
well in the metal-rich regime. However, in the metal-poor regime the RRc display a broader peak and steeper slope 
([Fe/H]$_{\Delta S} \lesssim$-1.8). This evidence indicates that the production rate of first overtone variables decreases as the metallicity
 of the stellar environment increases. This trend is expected because the metallicity is the most relevant parameter in shaping the
morphology of the horizontal branch \citep{2019A&A...629A..53T}, and in turn the sampling of the RRL instability strip 
\citep{1997A&AS..121..327B,2011rrls.conf....1B}.

\subsection{The new $\Delta$S calibration applied to a mixed mode variable}

We also note that we have applied the new calibration to a 
field mixed mode variable (ASAS~J183952-3200.9). On the basis of 10 high 
resolution spectra collected with X-Shooter and 2 with FEROS we estimated an 
iron abundance of [Fe/H]=-1.62$\pm$0.10. All 12 estimates were made from individual
spectra, i.e. without stacking. The [Fe/H]$_{\Delta S}$ abundance, 
based on the new $\Delta S$ calibration applied to the same spectra but 
degraded to a spectral resolution of R=2,000, agrees quite well with 
[Fe/H]$_{\Delta S}$=-1.51$\pm$0.12. This is the first 
time that the metallicity of an RRd is estimated by using $\Delta$S and the 
remarkable agreement with the direct measurement further supports the 
plausibility of a single $\Delta$S calibration for all RRLs regardless of 
their pulsation mode.


\section{Summary and final remarks} \label{ch:final_remarks}
We have provided a new calibration of the $\Delta$S method to derive low resolution spectroscopic metallicity estimates 
for RRL stars. It departs
directly from the $\Delta$S versus [Fe/H] plane and makes use of homogeneous HR metallicity measurements for 143 RRLs 
(111 RRab, 32 RRc) from a variety of spectrographs. The calibrating stars display a wide range of pulsational amplitudes
and periods. They cover over three dex in iron abundance, including the
metal-poor and metal-rich tails that were poorly sampled in previous works. The most metal-poor star in the CHR sample, 
with two high resolution measurements resulting in [Fe/H]=-3.06$\pm$0.08, may be among the most metal-poor RRLs ever identified 
\citep{2009ApJ...692L.127W,2011A&A...527A..65H}. The metal-rich tail is similarly remarkable, with multiple stars presenting
super-solar metallicities and pointing to a complex chemical enrichment history.

For the first time, this empirical calibration includes the full pulsation cycle 
as well as first overtone pulsators. Therefore, it is not necessary to classify 
the pulsation mode of the RRL of interest, nor execute timed observations in order 
to gather the data at specific phase intervals. This means no detailed knowledge 
of the light curve is required.

While we suggest preference be given to the calibration with all Balmer lines 
(H$_{\delta}$, H$_{\gamma}$, H$_{\beta}$) when possible, followed by combinations 
that include H$_{\beta}$, we also provide coefficients for all combinations of one 
or two of these lines. We have employed both SEGUE-SDSS (R$\approx$2,000) and LAMOST (R$\approx$1,500) 
spectra to investigate whether the new calibration is valid for low resolution spectra 
collected with different spectrographs. Our findings support that the new calibration
can be applied to different spectrographs and resolutions. This means that there is 
no need for transformations between different equivalent widths systems in order to
obtain metallicity estimates within the typical uncertainties of the method. We also 
provide 211 reference stars covering a wide range in metallicity in high resolution, 
144 of which 
have also [Fe/H]$_{\Delta S}$ estimates based on all combinations of Balmer lines. This 
is the largest sample of high resolution measurements for field RRL stars in the
literature and can be used to anchor any future works in our metallicity scale.

We have applied the new $\Delta$S calibration to a sample of 5,001 field RRLs (3,439 RRab, 1,562 RRc) with either SEGUE-SDSS 
low resolution spectra or high resolution spectra downgraded and rebinned in order to mimic the SEGUE-SDSS spectra. This resulted 
in a distribution with median $\eta$=-1.55$\pm$0.01 and $\sigma$=0.51, in good agreement with previous studies of halo
RRLs. For comparison, the HR sample described above has a distribution with $\eta$=-1.57$\pm$0.04 and $\sigma$=0.61.
In both cases, the distribution for the RRc alone peaks at slightly lower metallicities than the RRab. Furthermore,
the metallicity distributions of the RRab and RRc, when considered separately, have different profiles in the metal-poor
regime. Indeed, the slope of the distribution is shallower for the RRc. This difference in the profiles and the $\approx$0.1
difference in the peaks of the two distributions support the theoretical scenario of a steady decrease in the production 
of RRc as the metallicity of the stellar environment increases.

\acknowledgments

It is a real pleasure to thank the anynomous referee for her/his positive remarks concerning the content and the results of 
this investigation and for her/his pertinent suggestions that improved the readability of the paper.

This research has made use of the National Aeronautics and Space Administration (NASA) Astrophysics Data System,
the National Institute of Standards and Technology (NIST) Atomic Spectra Database, the JVO Portal\footnote{
\url{http://jvo.nao.ac.jp/portal/}} operated by ADC/NAOJ, and the ESO Science Archive Facility. 

Based on observations made with the Italian Telescopio Nazionale Galileo (TNG) operated on the island of La Palma 
by the Fundaci\'{o}n Galileo Galilei of the INAF (Istituto Nazionale di Astrofisica) at the Spanish Observatorio del 
Roque de los Muchachos of the Instituto de Astrofisica de Canarias.

Some of the observations reported in this paper were obtained with the Southern African Large Telescope (SALT).

Based on observations collected at the European Organisation for Astronomical Research in the Southern Hemisphere under ESO programmes 0100.D-0339, 0101.D-0697, 0102.D-0281, 076.B-0055, 077.B-0359, 077.D-0633, 079.A-9015, 079.D-0262, 079.D-0462, 079.D-0567, 082.C-0617,
083.B-0281, 083.C-0244, 094.B-0409, 095.B-0744, 097.A-9032, 098.D-0230, 189.B-0925, 267.C-5719, 297.D-5047, 67.D-0321, 67.D-0554, 69.C-0423, 71.C-0097, 0100.D-0273, 083.C-0244, 098.D-0230.

Guoshoujing Telescope (the Large Sky Area Multi-Object Fiber Spectroscopic Telescope LAMOST) is a National Major Scientific Project built by the Chinese Academy of Sciences. Funding for the project has been provided by the National Development and Reform Commission. LAMOST is operated and managed by the National Astronomical Observatories, Chinese Academy of Sciences.

Funding for the SDSS and SDSS-II has been provided by the Alfred P. Sloan Foundation,
the Participating Institutions, the National Science Foundation, the U.S. Department of
Energy, the National Aeronautics and Space Administration, the Japanese Monbukagakusho,
the Max Planck Society, and the Higher Education Funding Council for England. The SDSS
Web Site is \url{http://www.sdss.org/.}
The SDSS is managed by the Astrophysical Research Consortium for the Participating
Institutions. The Participating Institutions are the American Museum of Natural History,
Astrophysical Institute Potsdam, University of Basel, University of Cambridge, Case Western
Reserve University, University of Chicago, Drexel University, Fermilab, the Institute for Advanced Study, 
the Japan Participation Group, Johns Hopkins University, the Joint Institute
for Nuclear Astrophysics, the Kavli Institute for Particle Astrophysics and Cosmology, the
Korean Scientist Group, the Chinese Academy of Sciences (LAMOST), Los Alamos National
Laboratory, the Max-Planck-Institute for Astronomy (MPIA), the Max-Planck-Institute for
Astrophysics (MPA), New Mexico State University, Ohio State University, University of
Pittsburgh, University of Portsmouth, Princeton University, the United States Naval Observatory, 
and the University of Washington.

We acknowledge financial support from US NSF under Grants AST-1714534 (MM, JPM) and AST1616040 (CS). 
EKG, BL, and ZD were supported by the Deutsche Forschungsgemeinschaft (DFG, German Research Foundation)
- Project-ID 138713538 - SFB 881 ("The Milky Way System", subprojects A03, A05, A11). EV acknowledges 
the Excellence Cluster ORIGINS Funded by the Deutsche Forschungsgemeinschaft (DFG, German Research Foundation) 
under Germany's Excellence Strategy \-- EXC \-- 2094 \--390783311.


\appendix
\section{Equivalent width dependence on metallicity and phase} \label{ch:appendix_ew_feh_phase}

This work departed directly from the $\Delta$S versus spectroscopic metallicity plane in order to find the best 
analytical equation and parameters relating these two quantities. The original Layden94 equation, however, 
departed from a polynomial fit in the HK plane, resulting in a relation with the form

\begin{equation}
K = a + b H + c [Fe/H] + d H [Fe/H],
\end{equation}

\noindent where a to d are constants, K the EW of the Ca II K line and H the mean EW for the three H lines. This is a 
very reasonable strategy as there is a clear pattern for the RRL in this plane, with stars of similar metallicity 
clustering in a somewhat linear fashion, with a slope and intercept that increases with metallicity. Thus, we attempted 
a new fit of the coefficients of this equation. The result created a larger overall scatter in the $\Delta$S-[Fe/H]
plane with $\sigma_r$ greater by 0.12. Furthermore, it could not trace the metal rich regime. The polynomial equation 
derived in this work, on the other hand, does not contain a metallicity-dependent slope in the HK plane. This in turn
yielded a tighter relation that can reach the metal rich regime 
(Fig.~\ref{fig:compare_definitions_H_K_plane}). A comparison in the $\Delta$S-[Fe/H] plane between the polynomial fit 
adopted in this work and the Layden94 equation with new coefficients is shown in Fig.~\ref{fig:compare_equations_dS_FeH_plane}.

\begin{figure*}
\centering
\includegraphics[width=0.8\textwidth]{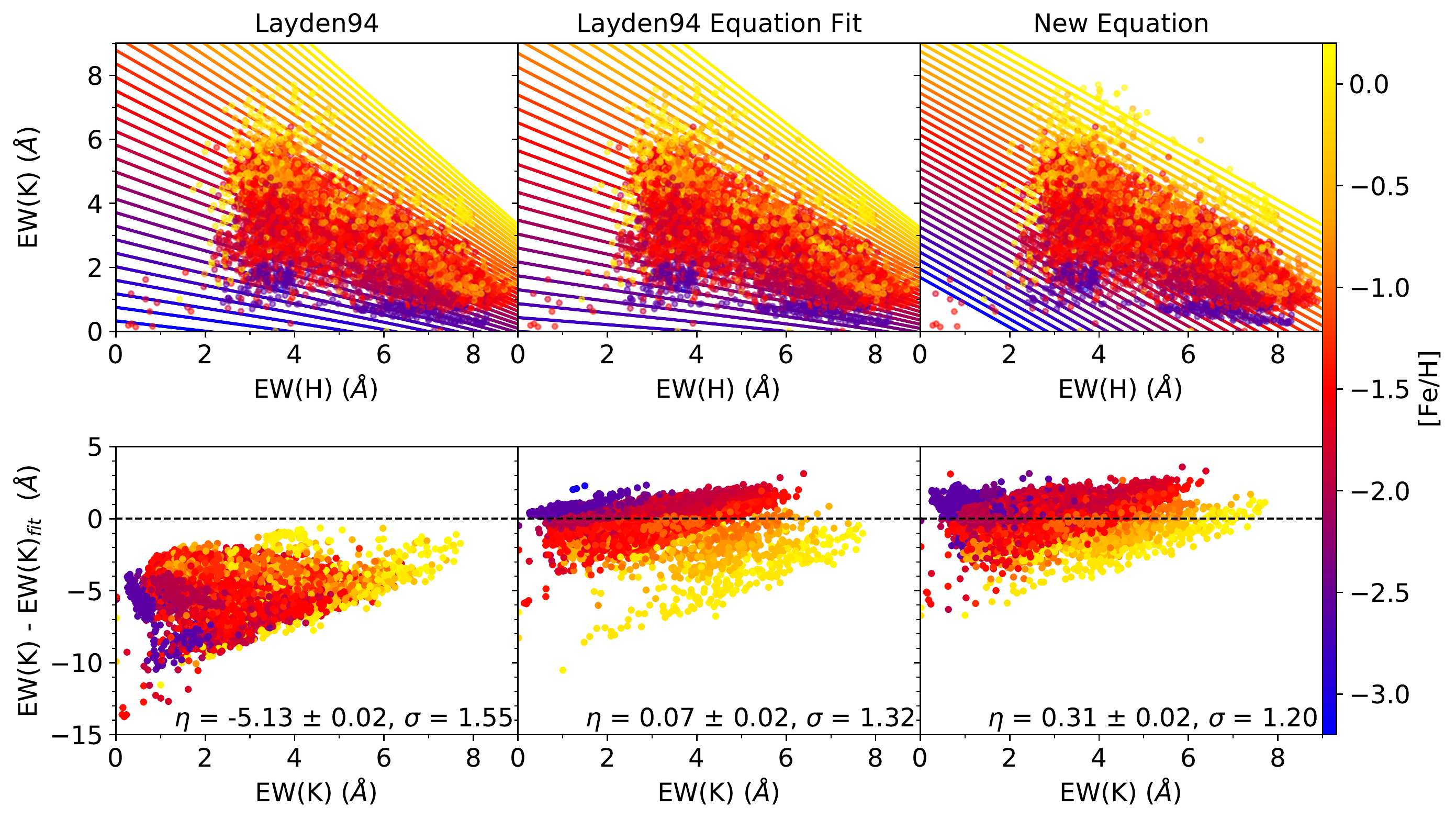}
\caption{The HK plane (top) and residuals of the $\Delta$S (bottom) for the CHR sample considering three cases: the 
original Layden94 equation (left), a
fit of the Layden94 equation with new coefficients (middle), and the new equation (right). The points are the measured EW values
in the new definition of wavelength ranges. The lines in the top panels are the predicted Ca II K equivalent width considering
each equation for a range of metallicity values. Data points and analytical lines are colored by high resolution metallicity 
according to the colorbar at the right edge of the figure. 
\label{fig:compare_equations_H_K_plane}}
\end{figure*}

The original Layden94 definition did not include rising branch phases due to the shock waves that occur at this point of
the pulsation cycle and can distort the profile of the Balmer lines. This phase avoidance introduces the burdensome 
necessity of properly phasing the RRL and timing observations in such a way as to avoid these phases. Adequate light curves 
and reference epochs require multiple previous observations that may not be readily available. Furthermore, as discussed 
previously, the duration of the line-distorting shock waves is very brief. 

\begin{figure*}
\centering
\includegraphics[width=0.6\textwidth]{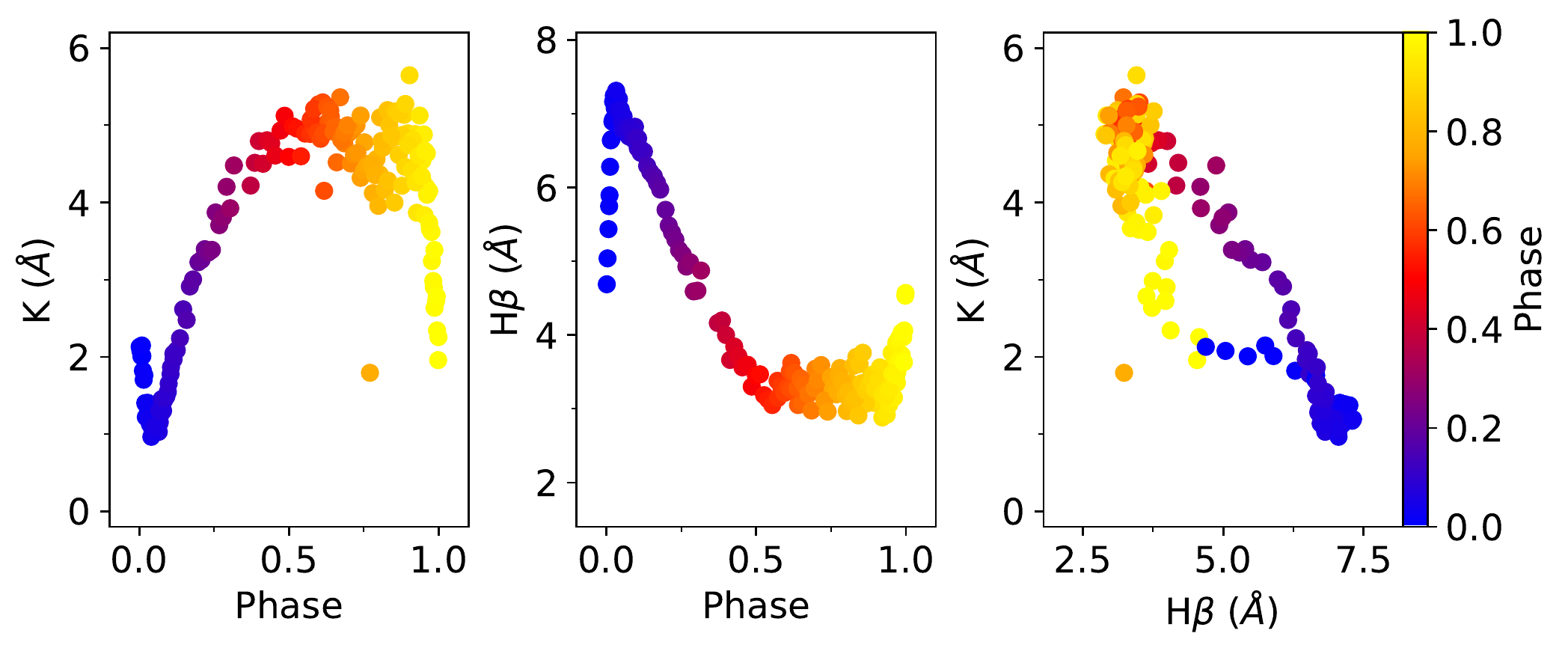}
\caption{Equivalent widths for the Ca II K (left) and H$\beta$ (middle) lines, and the HK plane (right) for HH Pup.
Symbols are colored according to phase. 
\label{fig:K_H_loop_analysis}}
\end{figure*}

\begin{figure*}
\centering
\includegraphics[width=0.7\textwidth]{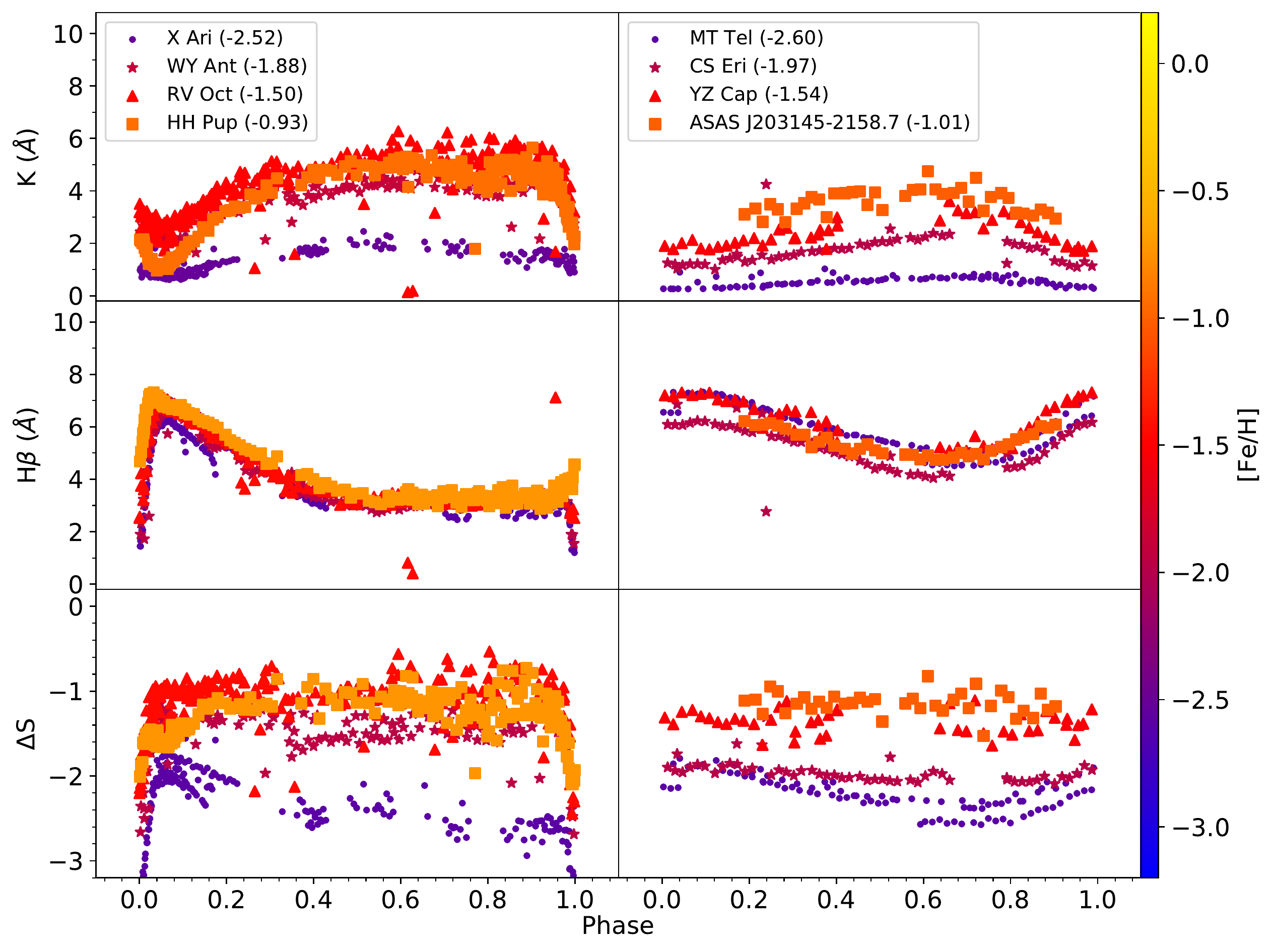}
\caption{Behavior of the equivalent width of the Ca II K line (top), H$\beta$ line (middle), and $\Delta $S value 
(bottom) across the entire pulsation cycle for three RRab (left panels) and three RRc (right panels). The points are 
colored by metallicity according to the colorbar on the right edge of the figure. The considered stars are listed in
the legend, with their respective high resolution metallicities between parentheses.
\label{fig:K_H_dS_versus_phase}}
\end{figure*}

\begin{figure*}
\centering
\includegraphics[width=0.65\textwidth]{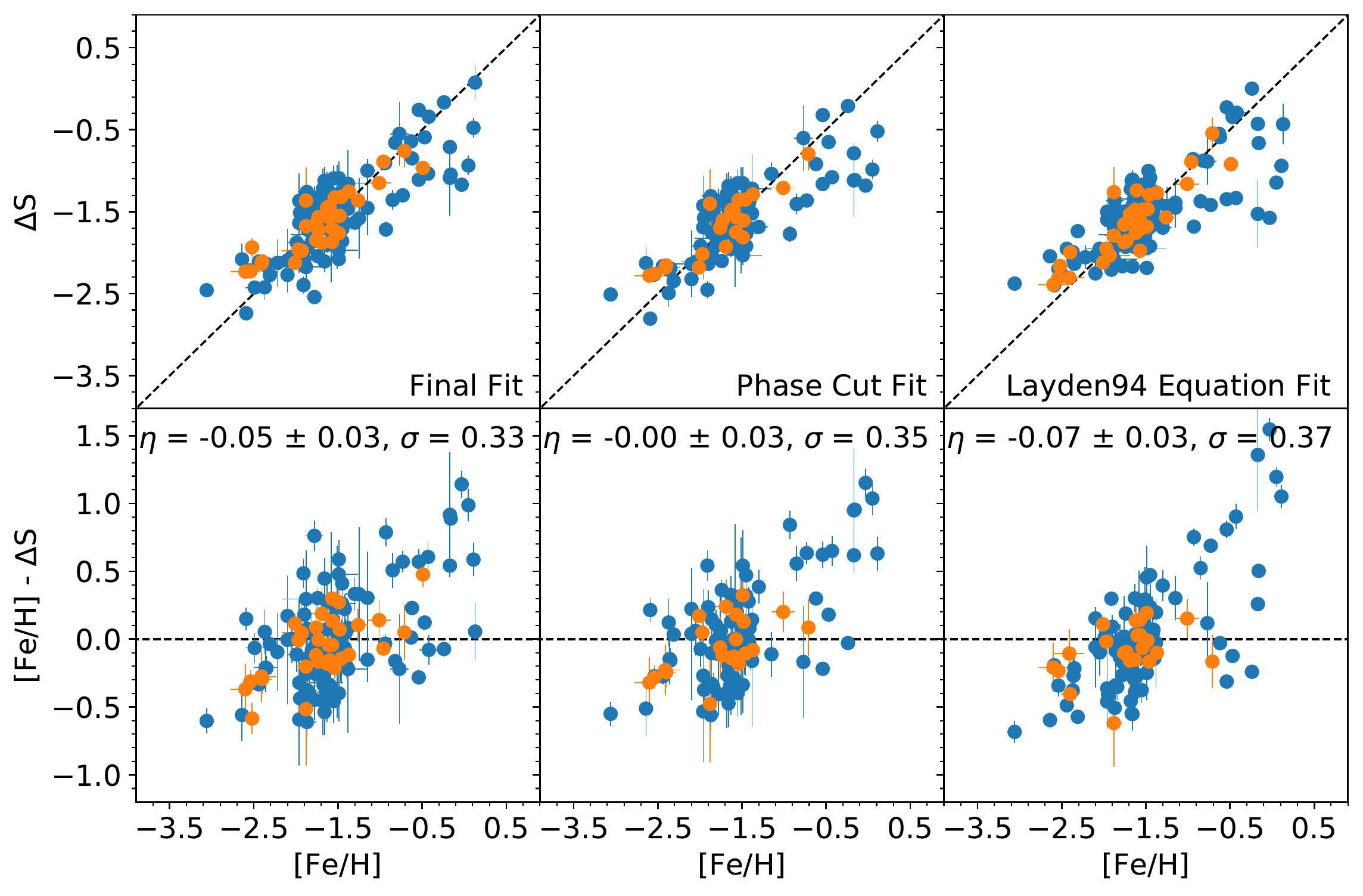}
\caption{The $\Delta$S-[Fe/H] plane (top) and residuals (bottom) for three cases considering all H lines: the final polynomial 
fit adopted in this work (left), the polynomial fit derived from the sample with phases between 0.1 and 0.9 (middle), and the fit
of the Layden94 equation with new coefficients (right). The RRab are plotted in blue and the RRc in orange.
\label{fig:compare_equations_dS_FeH_plane}}
\end{figure*}

In the HK plane, stars in these phases occupy a region underneath the area created by the other phases, effectively creating a
"loop" that invades a lower metallicity region. Fig.~\ref{fig:K_H_loop_analysis} illustrates the origin of the "loops". Their 
lower sequence is caused by not only a sharp peak in the EW of the H lines, but also a sudden dip in the EW of the Ca II K
line. Both the H and the Ca II K lines are strongly responsive to the effective temperature, albeit in opposite directions,
i.e. the EW of H lines increases with temperature, while that of the Ca II K decreases. Deriving metallicities from a combination
of these lines, therefore, relies on this mirrored effect that nullifies the influence of the temperature on the equivalent 
width of the Ca II K, effectively isolating the metallicity impact. However, the peak of the H lines and the dip of the Ca II K 
lines in EW are not perfectly synchronized, with the latter occuring slightly earlier. Thus, at phase 0.9 to 0.1, we observe a 
sudden drop of any value that relates these two quantities, as evident in the bottom panels of Fig.~\ref{fig:K_H_dS_versus_phase}.

Note that the EW variation of the lines of interest in the RRc is significantly smoother. This is also true for their
light curves, which are almost sinusoidal while those of their fundamental mode counterparts are sawtooth-shaped. The Ca II K 
line dip is not identical among RRab stars either. There is some evidence that more metal rich stars will display a deeper dip,
but other characteristics such as pulsational amplitude may play a role.

In order to investigate the effect of phases between 0.9 and 0.1 in the final $\Delta$S value, we have removed them from the 
sample and performed a new fit of the coefficients of the polynomial equation and of the Layden94 equation. We found that this 
produced slightly smaller median residuals and slightly larger scatter for the new equation 
(Fig.~\ref{fig:compare_equations_dS_FeH_plane}). For the Layden94 equation, the phase cut caused both quantities to increas over 
0.2. This phase interval, therefore, only has minimal effect on the $\Delta$S computation with the new equation.

\section{Supplementary tables}

\begin{deluxetable*}{lrrrrr} 
\tablecaption{Individual measurements of the equivalent widths of interest for the CHR sample. 
\label{tab:deltas_ews}}
\tablewidth{\textwidth}
\tablehead{
\colhead{GaiaID} & \colhead{MJD} & 
\colhead{CaII K} & \colhead{H$_{\delta}$} & \colhead{H$_{\gamma}$} & \colhead{H$_{\beta}$} \\
\colhead{(DR2)} & \colhead{(day)} & \colhead{(\AA)} & \colhead{(\AA)} & \colhead{(\AA)} & \colhead{(\AA)}
}
\startdata
15489408711727488 & 2455454.81452 & 2.14$\pm$0.57 & 4.05$\pm$0.62 & 4.26$\pm$0.38 & 3.24$\pm$0.35 \\
15489408711727488 & 2455454.82015 & 1.95$\pm$0.73 & 3.97$\pm$0.46 & 4.34$\pm$0.52 & 3.22$\pm$0.33 \\
15489408711727488 & 2455454.82434 & 1.94$\pm$0.55 & 4.26$\pm$0.46 & 4.08$\pm$0.48 & 3.26$\pm$0.54 \\
15489408711727488 & 2455454.85236 & 1.89$\pm$0.85 & 3.86$\pm$0.48 & 4.18$\pm$0.70 & 3.09$\pm$0.40 \\
15489408711727488 & 2455454.85655 & 2.09$\pm$0.72 & 3.67$\pm$0.55 & 3.87$\pm$0.58 & 2.96$\pm$0.41 \\
\enddata
\tablecomments{Table \ref{tab:deltas_ews} is published in its entirety in the machine-readable format.
A portion is shown here for guidance regarding its form and content.}
\end{deluxetable*}

\bibliography{bibliography_ds}{}
\bibliographystyle{aasjournal}

\end{document}